\documentclass[letterpaper, reprint, showkeys, footinbib, prl, aps, superscriptaddress, longbibliography, nobalancelastpage]{revtex4-1}

\usepackage[pdftex, breaklinks=true]{hyperref}
\hypersetup{
    pdffitwindow=true,     
    pdfstartview={FitH},    
    pdfnewwindow=true,      
    colorlinks=true,       
    linkcolor=red,      
    citecolor=blue,        
    urlcolor=blue           
}
\usepackage[utf8]{inputenc}
\DeclareSymbolFont{slenderlargesymbols}{OMX}{ccex}{m}{n}
\DeclareMathSymbol{\prod}{\mathop}{slenderlargesymbols}{"51}
\usepackage{amsfonts, amssymb, amsmath}
\usepackage{amsthm}
\usepackage{mathtools}
\usepackage{braket} 
\usepackage{multirow}
\usepackage[T1]{fontenc}

\newtheoremstyle{cited}%
  {3pt}
  {3pt}
  {\itshape}
  {}
  {\bfseries}
  {.}
  {.5em}
  {\thmname{#1} \thmnumber{#2} \thmnote{\normalfont#3}}
\theoremstyle{cited}

\makeatletter

\newcommand\k@t[1]{{|{#1}\rangle}}
\makeatother
\DeclarePairedDelimiter{\abs}{\lvert}{\rvert}

\DeclareMathOperator{\tr}{\text{tr}}
\DeclareMathOperator{\EX}{\mathbb{E}}

\usepackage{xcolor}

\usepackage[normalem]{ulem}

\usepackage{tikz}
\usetikzlibrary{decorations.pathreplacing}
\makeatletter
\def\underbrace#1{%
   \@ifnextchar_{\tikz@@underbrace{#1}}{\tikz@@underbrace{#1}_{}}}
\def\tikz@@underbrace#1_#2{%
   \tikz[baseline=(a.base)] {\node[inner sep=2] (a) {\(#1\)};
   \draw[line cap=round,decorate,decoration={brace,amplitude=4pt}]
     (a.south east) -- node[pos=0.5,below,inner sep=7pt] {\(\scriptstyle #2\)} (a.south west);}}
 
\def\overbrace#1{%
   \@ifnextchar^{\tikz@@overbrace{#1}}{\tikz@@overbrace{#1}^{}}}
\def\tikz@@overbrace#1^#2{%
   \tikz[baseline=(a.base)] {\node[inner sep=2] (a) {\(#1\)};
   \draw[line cap=round,decorate,decoration={brace,amplitude=4pt}]
     (a.north west) -- node[pos=0.5,above,inner sep=7pt] {\(\scriptstyle #2\)} (a.north east);}}
\makeatother

\usepackage{siunitx}

\begin{document}
\title{Machine Learning for Continuous Quantum Error Correction on Superconducting~Qubits}

\author{Ian Convy}
\thanks{These two authors contributed equally\\\href{mailto:ian_convy@berkeley.edu}{ian\_convy@berkeley.edu}\\\href{mailto:haoran.liao@berkeley.edu}{haoran.liao@berkeley.edu}}
\affiliation{Department of Chemistry, University of California, Berkeley, CA 94720, USA}
\affiliation{Berkeley Quantum Information and Computation Center, University of California, Berkeley, CA 94720, USA}
\affiliation{Challenge Institute for Quantum Computation, University of California, Berkeley, CA  94720}

\author{Haoran Liao}
\thanks{These two authors contributed equally\\\href{mailto:ian_convy@berkeley.edu}{ian\_convy@berkeley.edu}\\\href{mailto:haoran.liao@berkeley.edu}{haoran.liao@berkeley.edu}}
\affiliation{Department of Physics, University of California, Berkeley, CA 94720, USA}
\affiliation{Berkeley Quantum Information and Computation Center, University of California, Berkeley, CA 94720, USA}
\affiliation{Challenge Institute for Quantum Computation, University of California, Berkeley, CA  94720}

\author{Song Zhang}
\affiliation{Department of Physics, University of California, Berkeley, CA 94720, USA}
\affiliation{Berkeley Quantum Information and Computation Center, University of California, Berkeley, CA 94720, USA}
\affiliation{Challenge Institute for Quantum Computation, University of California, Berkeley, CA  94720}

\author{Sahil Patel}
\affiliation{Department of Computer Science, University of California, Berkeley, CA 94720, USA}
\affiliation{Berkeley Quantum Information and Computation Center, University of California, Berkeley, CA 94720, USA}
\affiliation{Department of Physics, University of California, Berkeley, CA 94720, USA}
\affiliation{Challenge Institute for Quantum Computation, University of California, Berkeley, CA  94720}

\author{William~P.~Livingston}
\affiliation{Department of Physics, University of California, Berkeley, CA 94720, USA}
\affiliation{Center for Quantum Coherent Science, University of California, Berkeley, California 94720, USA}

\author{Ho Nam Nguyen}
\affiliation{Department of Physics, University of California, Berkeley, CA 94720, USA}
\affiliation{Berkeley Quantum Information and Computation Center, University of California, Berkeley, CA 94720, USA}
\affiliation{Challenge Institute for Quantum Computation, University of California, Berkeley, CA  94720}

\author{Irfan Siddiqi}
\affiliation{Department of Physics, University of California, Berkeley, CA 94720, USA}
\affiliation{Center for Quantum Coherent Science, University of California, Berkeley, California 94720, USA}

\author{K. Birgitta Whaley}
\affiliation{Department of Chemistry, University of California, Berkeley, CA 94720, USA}
\affiliation{Berkeley Quantum Information and Computation Center, University of California, Berkeley, CA 94720, USA}
\affiliation{Challenge Institute for Quantum Computation, University of California, Berkeley, CA  94720}
\affiliation{Center for Quantum Coherent Science, University of California, Berkeley, California 94720, USA}

\date{\today}

\begin{abstract}

Continuous quantum error correction has been found to have certain advantages over discrete quantum error correction, such as a reduction in hardware resources and the elimination of error mechanisms introduced by having entangling gates and ancilla qubits. We propose a machine learning algorithm for continuous quantum error correction that is based on the use of a recurrent neural network to identify bit-flip errors from continuous noisy syndrome measurements. The algorithm is designed to operate on measurement signals deviating from the ideal behavior in which the mean value corresponds to a code syndrome value and the measurement has white noise. We analyze continuous measurements taken from a superconducting architecture using three transmon qubits to identify three significant practical examples of non-ideal behavior, namely auto-correlation at temporal short lags, transient syndrome dynamics after each bit-flip, and drift in the steady-state syndrome values over the course of many experiments. Based on these real-world imperfections, we generate synthetic measurement signals from which to train the recurrent neural network, and then test its proficiency when implementing active error correction, comparing this with a traditional double threshold scheme and a discrete Bayesian classifier. The results show that our machine learning protocol is able to outperform the double threshold protocol across all tests, achieving a final state fidelity comparable to the discrete Bayesian classifier.

\end{abstract}

\maketitle

\section{Introduction}

The prevalence of errors acting upon quantum states, either as a result of imperfect quantum operations or decoherence arising from interactions with the environment, severely limits the implementation of 
quantum computation on physical qubits. A variety of methods have been proposed to suppress the frequency of these errors, such as dynamic decoupling~\cite{Viola_Knill_Lloyd_1999}, application of a penalty Hamiltonian~\cite{Bookatz_Farhi_Zhou_2015}, decoherence-free subspace encoding~\cite{Lidar_Chuang_Whaley_1998}, and near-optimal recovery based on process tomography~\cite{Barnum_Knill_2002, Beny_Oreshkov_2011}. In addition to these tools for error prevention, there exist many schemes for quantum error correction (QEC) that are able to return the system to its proper configuration after an error occurs~\cite{Lidar_Brun_2013}. The ability to correct errors rather just suppress them is vital to the development of fault-tolerant quantum computation~\cite{Preskill_1997}.

An essential feature of QEC is the measurement of certain error syndrome operators, which provides information about errors on the physical qubits without collapsing the logical quantum state. In the canonical approach, quantum error correction is conducted in a discrete manner, using quantum logic gates to transfer the qubit information to ancilla qubits and subsequently making projective measurements on these to extract the error syndromes. However, in contrast to this theoretical idealization of instantaneous projections of the quantum state, experimental implementation of such measurements inherently involves performing weak measurements over finite time intervals~\cite{jacobs_straightforward_2006}, with the dispersive readouts in superconducting qubit architectures constituting the prime example of this in today's quantum technologies~\cite{blais_2004, blais_circuit_2021, krantz_quantum_2019, divincenzo_multi-qubit_2013}. This has motivated the development of continuous quantum error correction (CQEC)~\cite{Ahn_Doherty_Landahl_2002, Ahn_Wiseman_Milburn_2003, Ahn_Wiseman_Jacobs_2004, Sarovar_2004, Oreshkov_Brun_2007, Chase_Landahl_Geremia_2008, atalaya_BaconShor2020,
mohseninia_2020, atalaya_continuous_2021, Livingston_2021}, where the error syndrome operators are measured weakly in strength and continuously in time.

CQEC operates by directly coupling the data qubits to continuous readout devices. This avoids the ancilla qubits and periodic entangling gates found in discrete QEC, reducing hardware resources. Additionally, the presence of these entangling gate sequences and ancillas introduces additional error mechanisms, occurring in-between entangling gates or on ancillas, that can cause logical errors \cite{mohseninia_2020, Livingston_2021}. On noisy quantum hardware, multiple rounds of entangling gates and ancilla readouts are required to accurately identify the system state \footnote{In discrete QEC, full syndromes measurements are performed multiple times before attempting to decode, often $\mathcal{O}(n)$ times for a length $n$ repetition code or surface code \cite{Preskill}. This reduces the impact of faulty entangling gates or ancillas.}. All of this is also avoided by measuring data qubits directly, as in CQEC.

In addition to quantum memory, CQEC naturally lends itself to modes of quantum computation involving continuous evolution under time-dependent Hamiltonians, such as adiabatic quantum computing~\cite{RevModPhys.90.015002} and quantum simulation~\cite{RevModPhys.86.153}. Given that the Hamiltonians considered generally do not commute with the error operators, the action of an error induces spurious Hamiltonian evolution within the corresponding error subspace until the error is ultimately diagnosed and corrected, resulting in the accrual of logical errors~\cite{atalaya_continuous_2021}. CQEC can effectively shorten the spurious evolution time in the error subspaces, and therefore increase the target state fidelity in quantum annealing.

Previous theoretical work on CQEC has focused primarily on measurement signals that behave in an idealized manner~\cite{mohseninia_2020, atalaya_continuous_2021, atalaya_BaconShor2020}, such that each sample is assumed to be i.i.d. Gaussian with a mean given by one of the syndrome eigenvalues. However, in real dispersive readout signals we observe a wide variety of ``imperfections'' caused by hardware limitations and post-processing effects, which can lead to more complicated syndrome dynamics or significant alterations to the noise distribution. A well-calibrated CQEC protocol should be designed to take into account any significant non-ideal behavior for a given architecture. However, it is often difficult to generate a precise mathematical description of the imperfections present in real measurement signals. 

Machine learning algorithms offer a solution to this problem, as they can be optimized to solve a task by looking directly at the relevant data instead of relying on hard-coded decision rules. Highly expressive models involving multiple neural network layers have proven to be particularly effective at solving complex tasks such as image recognition and language translation~\cite{Goodfellow_Bengio_Courville_2016}. The recurrent neural network (RNN) is a popular sequential learning model, because it operates on inputs of varying length and provides an output at each step.
After being trained on a set of non-ideal measurement signals, an RNN can function as a CQEC algorithm by generating probabilities which describe the likelihood of an error at a given time step. Most importantly, the flexibility of the algorithm allows it to handle imperfections in the signal that would otherwise be impractical to model.

In this paper we investigate the performance of an RNN-based CQEC algorithm which acts on measurement signals with non-ideal behavior. We emphasize here \textit{active correction}, in which errors are corrected during the experiment as soon as they are observed. To quantify the benefits of using a neural network, we compare the RNN to a conventional double threshold scheme as well as to a discrete Bayesian classifier. The first threshold scheme for CQEC was by Sarovar \textit{et al.}~\cite{Sarovar_2004}, who used the sign of the averaged measurement signals (i.e.,\ a threshold at zero) to identify the error subspace. This filter was improved upon in Atalaya \textit{et al.}~\cite{atalaya_BaconShor2020} and Atalaya, Zhang \textit{et al.}~\cite{atalaya_continuous_2021}, as well as in Mohseninia \textit{et al.}~\cite{mohseninia_2020}, by adding a second threshold to better detect errors that affect multiple syndromes. We chose to compare our RNN model to the threshold scheme in \cite{atalaya_continuous_2021}, since it had superior performance in numerical tests (see App.~\ref{app:boxcar}). 

The remainder of the paper is structured as follows. Sec.~\ref{sec:background} reviews the three-qubit bit-flip code that will be used to evaluate the three models, and outlines the idealized mathematical formulation of CQEC. In Sec.~\ref{sec:setup} we use physical experimental data to characterize the imperfections that are present in typical superconducting qubit signals. We find that the noise possesses a significant amount of auto-correlation, while the syndromes demonstrate complex transient behavior after every bit-flip, as well as drift of the mean values over time. Sec.~\ref{sec:models} then describes in detail the double threshold, discrete Bayesian, and RNN-based models that we will be comparing. In Sec.~\ref{sec:experiments} we test the error correction capabilities of the models using four different sets of synthetic data, each displaying a different characteristic feature or set of features of non-ideal behavior. We show that the RNN is able to outperform the double threshold across all synthetic experiments, achieving results comparable to those of the Bayesian model. Sec.~\ref{sec:discussion} summarizes our findings and proposes directions for future work. 

\section{Background}\label{sec:background}
We exemplify our CQEC protocol by operating it on the three-qubit bit-flip stabilizer code; in general, the protocol works with any QEC codes. The three-qubit bit-flip stabilizer code encodes the logical states $\ket{0}$ and $\ket{1}$ into $\ket{0}_\text{L}=\ket{000}$ and $\ket{1}_\text{L}=\ket{111}$, respectively, where the stabilizer generators are chosen to be $S_1=Z_1Z_2$ and $S_2=Z_2Z_3$, which also serve as the error syndrome operators. The states $\ket{000}$ and $\ket{111}$ span the code subspace, in which the syndromes have values $(S_1=+1, S_2=+1)$. The $(S_1=-1, S_2=+1)$, $(S_1=-1, S_2=-1)$, $(S_1=+1, S_2=-1)$ subspaces are known as the error subspaces, which are spanned by the basis states $\{\ket{011},\ket{100}\}$, $\{\ket{010},\ket{101}\}$ and $\{\ket{001},\ket{110}\}$, respectively. A logical error in quantum memory, i.e., when there is no Hamiltonian evolution, is an error attributed to the logical $X$ operator, $X_\text{L}=X_1X_2X_3$.

In the continuous operation of the three-qubit bit-flip code, the error syndrome operators $S_k, k=\{1,2\}$ are continuously and simultaneously measured to yield the following idealized signals for each $S_k$ as a function of time $t$:
\begin{equation}\label{eq:signal_expression}
    I_{k}(t) =\sqrt{\Gamma_m^k} \text{tr}[S_k \rho(t)] + \xi_{k}(t).
\end{equation}
Here $\rho(t)$ is the density matrix of the three physical qubits and
$\Gamma_m^k$ is the measurement strength that determines the time to sufficiently resolve the mean values of the syndromes under constant variance. Specifically, $1/\Gamma_m^k$ is the time needed to distinguish between the eigenvalues of $S_k$ with a signal-to-noise ratio (SNR) of 1~\footnote{The SNR is defined as $(\mu_e-\mu_o)^2/(\sigma_e+\sigma_o)^2$, where $\mu$ and $\sigma$ are the mean and standard deviation of the signals, and subscripts denote the odd and even parities of the syndrome measurements.}. In the Markovian approximation, $\xi_{k}(t)$ is Gaussian white noise, i.e., $\xi(t)=\dot{W}(t)$ where $W(t)$ is a Wiener process, with a two-time correlation function $\langle\xi_{k}(t)\xi_{k'}(t')\rangle=\delta_{kk'}\delta(t-t')$, where the $\langle\cdot\rangle$ denotes average over an ensemble of noise realizations. In the continuous operation, the observer receives noisy voltage traces with means proportional to the syndrome operator eigenvalues and variances that determine the continuous measurement collapse timescales. Monitoring both error syndromes with streams of noisy signals represents a gradual gain of knowledge of the measurement outcome to diagnose bit-flip errors that occur. We shall refer to the parity of $I_k(t)$ as even or odd depending on whether the mean value of $I_k(t)$ is positive or negative. 
In an actual experiment we will only have access to the averaged signals taken at discrete time steps separated by $\Delta t$, which we denote by $I_{k, t}$ at time step $t$:
\begin{equation}\label{eq:discrete_signal}
    I_{k,t}=\sqrt{\Gamma_m^k}\tr\left[S_k\rho(t)\right] +\frac{\Delta W}{\Delta t}
\end{equation}
where $\Delta W\sim\mathcal{N}(0,\Delta t)$. We shall assume that $\rho(t)$ only changes due to bit-flips at the beginning of each time step $\Delta t$ for very small $\Delta t$. 

In previous work, Ref.~\cite{mohseninia_2020} compared the performance of a linear approximate Bayesian classifier and the double threshold model with one threshold fixed at $y=0$ and another threshold at $y>0$ in correcting the three-qubit bit-flip code for quantum memory. Ref.~\cite{atalaya_continuous_2021} analyzed the double threshold model with two varying thresholds in correcting the three-qubit bit-flip code, and applied it to quantum annealing under bit-flip errors $X_q$ with which the chosen annealing Hamiltonian does not commute. In the current work, we shall study the performance of machine learning algorithms both in quantum memory and in quantum annealing.

The stochastic master equation (SME)~\cite{jacobs_straightforward_2006} governing the evolution of $\rho(t)$ under measurements with a finite rate of information extraction implied by Eq.~\eqref{eq:signal_expression} in the presence of bit-flip errors is given by \cite{Sarovar_2004,atalaya_continuous_2021}
\begin{widetext}
\begin{equation}\label{eq:SME}
\dot{\rho}(t)=-i[H(t), \rho]+
\sum_{k=1,2}\left[\frac{\Gamma_\phi^k}{2}\left(S_{k} \rho S_{k}-\rho\right)+\sqrt{\Gamma_m^k}\xi_k(t)\left(\frac{S_{k} \rho+\rho S_{k}}{2}-\rho \braket{S_{k}}_\rho\right)\right]
+\sum_{q=1,2,3} \gamma_{q}\left(X_{q} \rho X_{q}-\rho\right).
\end{equation}
\end{widetext}
The first term describes coherent evolution of the three-qubit state under a Hamiltonian $H(t)$, which can, for instance, be a quantum annealing Hamiltonian. The second term describes the back-action induced by the simultaneous continuous measurement of the error syndrome operators $S_1$ and $S_2$ on the three-qubit state, where $\Gamma_\phi^k$ is the measurement-induced ensemble dephasing rate of the corresponding error syndrome operator $S_k$. The measurement strength $\Gamma_m^k$, is related to the detector efficiency $\eta_k$ as $\Gamma_m^k=2\Gamma_\phi^k\eta_k$
The first two terms can be obtained by substituting operators $c_k\propto S_k$ into the general SME $d\rho=-i[H, \rho]dt+\sum_k(\mathcal{D}[c_k]\rho dt+\sqrt{\eta_k}\mathcal{H}[c_k]\rho dW)$. The third term describes the decoherence of the three-qubit state in the presence of bit-flip errors, with $\gamma_q, q=\{1,2,3\}$ denoting the bit-flip error rate of the $q^\text{th}$ physical qubit. While the idealized measurement signals mentioned above assume no effect induced by physical experimental apparatus in the qubit readouts, there are various imperfections of the measurement signals in practice that make the error diagnosis more challenging. We shall first present the characteristics of these measurement signals from physical experiments below and explain their implications for our purpose. 

\section{Problem Setup}\label{sec:setup}

\begin{figure}
 \centering
\includegraphics[scale=0.6]{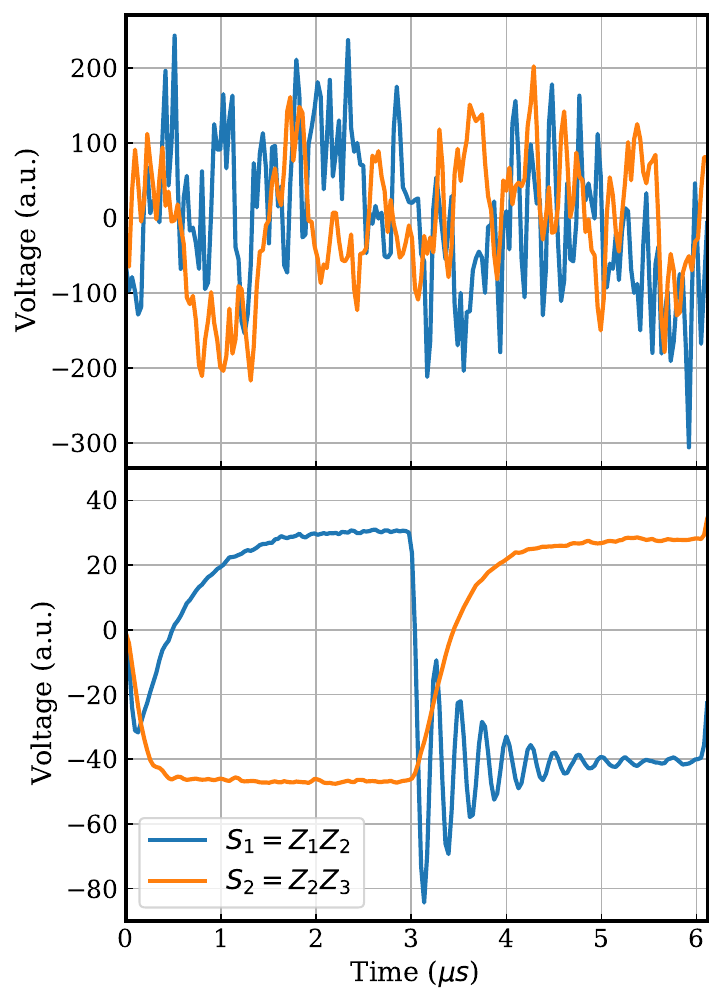}
 \caption{The measurement signals of the two syndrome operators $S_1=Z_1Z_2$ and $S_2=Z_2Z_3$ on the transmon qubits. The even(odd) parity signal, i.e., $S_k=+1$($-1$) has a voltage readout that is centered at an arbitrary negative(positive) value, according to Eq.~\eqref{eq:voltage}. We note that the experimental voltage readout of even parity is centered at the negative mean by design. The upper figure is the raw voltage signal readout of a single experimental run. The lower figure is the averaged voltage readout over $47,494$ post-selected runs. The qubits are initialized to $\ket{100}$ and an $X_2$ bit-flip is artificially injected at $t=\SI{3.0}{\micro \second}$, resulting in a new state $\ket{110}$. The oscillation pattern is explained in App.~\ref{app:transients}.}
 \label{fig:signal_example}
\end{figure}

\subsection{Characteristics of CQEC Measurement Signals}

The superconducting qubits are monitored using voltage signals from homodyne measurements of the parity operators that are derived from tones reflected off the resonator (see App.~\ref{app:homodyne}). The resonator signal is fed into a Josephson parametric amplifier (JPA) in order to increase the signal strength without adding a significant amount of noise. The amplified radio frequency signals are then demodulated and digitized. After a further digital demodulation, the signals are processed with an exponential anti-aliasing filter with a time constant of \SI{32}{\nano\second}. This filtered signal, which is averaged in $\Delta t = \SI{32}{\nano\second}$ bins, is then streamed from the digitizer card to the computer.

Due to the effects of the amplifier and resonator, we expect that measurements performed on such real superconducting devices will deviate from the idealized behavior predicted by Eq.~\eqref{eq:signal_expression}. In particular, we can anticipate the following three imperfections: 
\begin{enumerate}
    \item The noise will possess a high degree of positive auto-correlation at short temporal lags due to the narrow low-pass bandwidth of the JPA and anti-aliasing filter.
    
    \item When a bit-flip occurs, the syndrome means will change gradually rather than instantaneously as the resonator reaches its new steady state. These periods are referred to as \textit{resonator transients} to stress their temporary nature, and arise because of time-dependent changes in the measurement strength $\Gamma^k_m$ (see App.~\ref{app:transients}).
    
    \item The values of the syndromes will drift over time due to small changes in experimental conditions (e.g. temperature). Unlike the other imperfections, this effect is only noticeable when comparing \textit{across} quantum trajectories rather than within them.
\end{enumerate}
These non-ideal behaviors in the measurement signals extracted from our typical physical experiments will be incorporated into our simulated experiments in Sec.~\ref{sec:experiments}.

Fig.~\ref{fig:signal_example} shows experimental dispersive readouts taken from three transmon qubits~\cite{Koch_transmon_2007} over the span of \SI{6}{\micro\second}~\cite{Livingston_2021}. The blue and orange lines are a record of the outputs from the two resonators, each measuring a different pair of qubits for their syndromes. The top figure shows the measurement signals from a single experiment, which contain large amounts of auto-correlated noise. During the experiment an $X_2$ error was injected at \SI{3.0}{\micro\second}, flipping the system from $\ket{100}$ to $\ket{110}$, but the weak-measurement noise largely obscures its effect on the syndrome values. 

To reveal these underlying syndromes, the bottom figure of Fig.~\ref{fig:signal_example} shows an average over the measurements from roughly 47,500 experiments, each initialized to $\ket{100}$ and injected with an $X_2$ error at \SI{3.0}{\micro\second}. It takes approximately \SI{2}{\micro\second} after initialization for the syndromes to reach their steady-state values for $\ket{100}$, as the number of photons in each resonator increases from zero gradually. We ignore this effect in our analysis, as it will only occur once at the start of an experiment. After the $X_2$ error is injected, the syndromes do not instantaneously jump to a new pair of values but instead enter a transitory period which can include significant oscillations. These transients derive from the time-dependent changes in the measurement rate $\Gamma^k_m(t)$ analyzed in App.~\ref{app:transients}. This period lasts for roughly \SI{2}{\micro\second}, after which the syndromes stabilize at their new steady-state values for $\ket{110}$. 

Depending on the underlying hardware, a measurement signal may be generated on a wide variety of different scales, such as the arbitrary voltage scale in Fig.~\ref{fig:signal_example}. To denote a signal generically on any scale, we write the measurement samples as
\begin{equation}\label{eq:scaled_signal}
    I_{k, t} = \bar{S}_{k, t} + \sqrt{\tau_k}\varepsilon_t,
\end{equation}
where $\bar{S}_{k, t}$ is the scaled mean of the $k$-th resonator at step $t$, $\tau_k$ is the scaled variance of the $k$-th resonator, and $\varepsilon_t \sim \mathcal{N}(0, 1)$. In this notation, the physical quantities $\Gamma_m$ and $\Delta t$ from Eq.~\eqref{eq:discrete_signal} have been absorbed into $\bar{S}_{k, t}$ and $\tau_k$.

\subsection{Impact of Auto-correlations}\label{sec:autocorrelation}

Unlike the other imperfections, the challenge posed by auto-correlated signal noise can be characterized theoretically. If the Gaussian noise in $I_{k,t}$ is correlated, then the distribution of noise samples can be parameterized in terms of a covariance matrix $\Sigma$ whose off-diagonal elements determine the degree of correlation. For simplicity we restrict our analysis to dependencies that are Markovian, such that $I_{k, t}$ depends only on the preceding measurement $I_{k, t-1}$, though our conclusions are not limited to this regime. Using a correlation coefficient of $0 < \rho <1$, the joint Gaussian log-density describing $I_{k, t}$ and $I_{k, t-1}$ is
\begin{equation*}
\begin{split}
    \log&\ p(I_{k, t}I_{k, t-1}|\bar{S}_{k,t}) =
    \\
    &-\frac{1}{2\tau_k(1-\rho^2)}
    \begin{bmatrix}
        \tilde{I}_{k, t} &  \tilde{I}_{k, t-1}
    \end{bmatrix}
    \begin{bmatrix}
        1 & -\rho \\
        -\rho & 1
    \end{bmatrix}
    \begin{bmatrix}
         \tilde{I}_{k, t} \\  \tilde{I}_{k, t-1}
    \end{bmatrix} + A,
\end{split}
\end{equation*}
where $ \tilde{I}_{k, j} \equiv I_{k, j} - \bar{S}_{k, j}$ denotes the centered signal sample at step $j$ and $A$ is the log of the normalization constant. We shall assume hereafter that the signal has been rescaled such that $\bar{S}_{k, j} = \pm 1$.

The effect of auto-correlations on error correction is best characterized in terms of how it impacts the usefulness of the syndrome measurements. To be more precise, we know that the purpose of each measurement is to provide some information about whether the underlying syndrome value of the state is $1$ or $-1$. When framed in these terms, we can formalize and quantify a notion of measurement ``usefulness'' using Bayesian theory, specifically a ratio called the \textit{Bayes factor} which we denote as $\phi$~\cite{Kass_Raftery_1995}. This factor can be written in log form as
\begin{equation}\label{eq:bayes_factor}
\begin{split}
 &\log\phi_{k, t} = 
 \\
 &\log p(I_{k, t}|I_{k, t-1}, \bar{S}_{k, t} = 1) - \log p(I_{k, t}|I_{k, t-1}, \bar{S}_{k,t} = -1),
\end{split}
\end{equation}
and quantifies how much evidence $I_{k, t}$ gives about the underlying syndrome value if we have already seen the previous measurement $I_{k, t-1}$. The larger the magnitude of $\log \phi_{k, t}$ the more useful $I_{k, t}$ is for our task, with its sign simply indicating whether the evidence supports a value of $1$ or $-1$.

Let $Q=\Sigma^{-1}$. By making the substitutions $\sigma^{-1}=Q_{22}$ and $\mu=\bar{S}_{k, t}-Q_{12}/Q_{22}(I_{k,t-1}-\bar{S}_{k, t})$ in the unconditional log-densities $-(I_{k, t}-\mu)^2/(2\sigma)+A$, each of the conditional log-densities in Eq.~\eqref{eq:bayes_factor} can be written as
\begin{equation*}
\begin{split}
    \log p(I_{k, t}|I_{k, t-1}, &\bar{S}_{k, t}) =
    \\
    &-\frac{[I_{k, t} - \bar{S}_{k, t} -\rho(I_{k, t-1} - \bar{S}_{k, t})]^2}{2\tau_k(1- \rho^2)} + A,
\end{split}
\end{equation*}
where $A$ is again the normalization constant~\cite{Rue_Held_Held_2005}.
Expanding the numerator and keeping only the terms that depend on $\bar{S}_{k, t}$ gives
\begin{equation*}
    \log p(I_{k, t}|I_{k, t-1}, \bar{S}_{k, t}) \rightarrow \frac{S^2_k(\rho - 1) + 2\bar{S}_{k, t}(I_{k, t} - \rho I_{k, t-1})}{2\tau_k(1 + \rho)},
\end{equation*}
where we ignore the other terms since they will cancel when computing $\log \phi_{k, t}$. After substituting this representation back into Eq.~\eqref{eq:bayes_factor} we get
\begin{equation}\label{eq:bayes_factor_subs}
    \log \phi_{k, t} = \frac{2(I_{k, t} - \rho I_{k, t-1})}{\tau_k(1 + \rho)},
\end{equation}
where the value of $\log \phi_{k, t}$ depends not only on $I_{k, t}$ and $I_{k, t-1}$ but also on the variance and auto-correlation of the measurements.

To see the impact of the auto-covariance more clearly, we compute the expectation value $\EX[\log\phi_{k, t}]$ with respect to a Gaussian distribution centered on the true syndrome value $S'_{k,t}=\pm 1$. Since Eq.~\eqref{eq:bayes_factor_subs} is linear, we can simply substitute in $S'_{k,t}$ for $I_{k, t}$ and $I_{k, t-1}$ to get $\EX[\log\phi_{k, t}]$. After taking its magnitude, we have
\begin{equation}\label{eq:avg_bayes_factor}
    \abs{\EX[\log \phi_{k, t}]} = \frac{2(1 - \rho)}{\tau_k(1 + \rho)},
\end{equation}
which decreases as the value of $\rho$ increases. Eq.~\eqref{eq:avg_bayes_factor} shows that positive auto-correlation $(\rho > 0)$ in the signal makes each of our measurements less useful than if the noise had been uncorrelated ($\rho = 0$), which means that it will take longer for us to determine the value of $\bar{S}_{k, t}$ at a given measurement strength.

This result can be understood by imagining that $\bar{S}_{k, t}$ and $I_{k, t-1}$ are competing to determine the value of $I_{k, t}$, with smaller $\rho$ favoring $\bar{S}_{k, t}$. The more that $\bar{S}_{k, t}$ affects the measurement, the more that the measurement in turn tells us about $\bar{S}_{k, t}$ and thus the more useful it is to us. When $\rho$ is large, the value of $I_{k, t}$ tends to lie very close to the value of $I_{k, t-1}$ regardless of whether $\bar{S}_{k, t}$ is $1$ or $-1$, and therefore the measurement does not reveal much new information about the syndrome.

\section{Models}\label{sec:models}
\subsection{Double Thresholds}\label{sec:threshold}
The double threshold protocol from \cite{atalaya_continuous_2021} uses two standard signal processing methods, filtering and thresholding, to identify errors. The raw measurement signal is first passed through an exponential filter to smooth out oscillations, and then this averaged value is compared to a pair of adjustable threshold values to determine the state of the system. A slightly different double threshold protocol was proposed in \cite{mohseninia_2020}, which used boxcar averaging and fixed one of the thresholds at zero.

To estimate the definite error syndromes from the noisy measurements, we first filter the raw signals $I_k(t)$ to obtain corresponding filtered signals $\mathcal{I}_{k}(t)$ according to
\begin{equation*}
   \dot{\mathcal{I}}_{k}(t)=-\frac{\mathcal{I}_{k}(t)}{\tau}+\frac{I_{k}(t)}{\tau},
\end{equation*}
where $\tau$ is the averaging time parameter, and whose discretized version is similar. In the regime where $t-t_0\gg\tau$ where $t_0$ is at the last filtered signal reset, $\mathcal{I}_{k}(t)$ reads as
\begin{equation*}
    \mathcal{I}_{k}(t)=\int_{t_0}^{t} d t^{\prime} \frac{e^{-\frac{t-t^{\prime}}{\tau}}}{\tau} I_{k}\left(t^{\prime}\right).
\end{equation*}

\subsubsection{Thresholds for Error Correction}
After filtering the measurement signals, we then apply a double thresholding protocol to the filtered signals $\mathcal{I}_1(t)$ and $\mathcal{I}_2(t)$ that is parameterized by the two thresholds $\Theta_1$ and $\Theta_2$, where $\Theta_1$ is the threshold for the $-1$ value of the error syndromes and $\Theta_2$ is the threshold for the $+1$ value of the error syndromes. If at least one of $\mathcal{I}_1(t)$ or $\mathcal{I}_2(t)$ is found to lie within the interval $(\Theta_1, \Theta_2)$, we declare to be uncertain of the error syndromes and do not perform any error correction operation. Otherwise, we apply the following procedure, in accordance with the standard approach for error diagnosis and correction. If both $\mathcal{I}_1(t) > \Theta_2$ and $\mathcal{I}_2(t) > \Theta_2$, then we diagnose the error syndromes as $(S_1 = +1, S_2 = +1)$ and accordingly perform no error correction operation. If $\mathcal{I}_1(t) < \Theta_1$ and $\mathcal{I}_2(t) > \Theta_2$, then we diagnose the error syndromes as $(S_1 = -1, S_2 = +1)$ and accordingly perform the error correction operation $C_\text{op} = X_1$. If both $\mathcal{I}_1(t) < \Theta_1$ and $\mathcal{I}_2(t) < \Theta_1$, then we diagnose the error syndromes as $(S_1 = -1, S_2 = -1)$ and accordingly perform the error correction operation $C_\text{op} = X_2$. If $\mathcal{I}_1(t) > \Theta_2$ and $\mathcal{I}_2(t) < \Theta_1$, then we diagnose the error syndromes as $(S_1 = +1, S_2 = -1)$ and accordingly perform the error correction operation $C_\text{op} = X_3$.

In quantum annealing, we note that the error correction operations are applied immediately after the error syndromes are diagnosed to minimize the aforementioned spurious Hamiltonian evolution. The action of an error correction operation $C_\text{op}$, assumed to be instantaneous, changes the three-qubit state $\rho(t)$ according to
\begin{equation*}
    \rho(t) \rightarrow C_\text{op}\rho(t)C_\text{op},
\end{equation*}
which applies to other models in our work as well. We note that the parameters $\{\tau, \Theta_1, \Theta_2\}$ constitutes the minimal set of tunable parameters. When the measurement signals $I_k$ have white noise, their optimal values in minimizing the logical error rate can be obtained by Eq.~(43) in~\cite{atalaya_continuous_2021} together with numerical optimizations.

We further reset the filtered signals $\mathcal{I}_k(t)$ to the corresponding initial syndrome value, at the same instant to avoid the transient delay in the filtered signals to reflect the application of the error correction operation on the state. Inherent within any error correction protocol, however, is the implicit assumption that the correction properly removes the error, which may not necessarily be the case if the error was misdiagnosed. 

We note that the $\mathcal{I}_k(t)$ used by the double threshold model in CQEC consists of weighted contributions from every raw signal taken prior to $t$ and after the last correction. The discrete Bayesian model and the RNN-based model that we discuss in this work can both be operated on raw signals, using all historical signals taken prior to a given $t$. This is in contrast to the projective measurement on ancilla superconducting qubits in discrete QEC that applies a matched filter \cite{PhysRevA.91.022118} on raw signals taken only within each detection round.

\subsection{Discrete Bayesian Classifier}\label{sec:bayesian}
One weakness of the double-threshold scheme is that its predictions are essentially all-or-nothing, since there is no in-built quantity that expresses the model's confidence. This contrasts with probabilistic classifiers, which generate probability values for each prediction class instead of only a single guess. By framing the classification problem in terms of probabilities, we can incorporate our knowledge of the error and noise distributions into our model in a mathematically rigorous manner.

Since each qubit in our system will experience either one or zero net flips after every time step, there are eight different ways that a state can be altered by bit-flips and therefore eight different classes that our classifier must track. We denote each of the possible bit-flip configuration using the state that $\ket{000}$ is taken to by the error, such that $\ket{001}$ denotes a flip on the third qubit, $\ket{110}$ denotes a flip on the first and second qubits, and so on. The goal of a probabilistic error corrector is to accurately determine the probability of all eight ``error states'' at time step $t$ given the measurement histories $\mathcal{M}^k_t \equiv \{I_{k,t'}\}_{t'=0}^{t'=t}$. We write this posterior probability as
\begin{equation}\label{eq:posterior_prob}
    \hat{p}(s_t) \equiv p(s_t|\mathcal{M}^1_t\mathcal{M}^2_t),
\end{equation}
where $s_t \in \{0, ..., 7\}$ denotes the digital representation of the error state at step $t$.

In the remainder of this subsection we consider a probabilistic classifier constructed using Bayes' theorem, which makes prediction based on the posterior probabilities of the different basis states at each time step~\cite{Sivia_Skilling_2006}. Starting with the knowledge of the initial state, this model uses a Markov chain and a set of Gaussian likelihoods to update our beliefs about the system conditioned on the specific measurement values that we observe. 

The Bayesian algorithm described in this section is derived by assuming that the mean of a given measurement $I_{k, t}$ is always determined by the state of the system at the end of the time step. This is equivalent to assuming that errors always happen at the beginning of each time step (see Sec.~\ref{sec:background}). Since our method for generating quantum trajectories follows this assumption, the Bayesian model is theoretically optimal for the numerical tests carried out in Sec.~\ref{sec:experiments} without mean drift or resonator transients. As the length of the step $\Delta t$ between measurements goes to zero, this algorithm converges to the Wonham filter~\cite{Wonham_1964}, which is known to be optimal for continuous quantum filtering of error syndromes~\cite{Mabuchi_2009}. This filter is similar to the discretized, linear Wonham filter derived in \cite{mohseninia_2020}, except that our filter does not rely on first-order approximations of the Markov evolution or Gaussian functions.

\subsubsection{Model Structure}

Using Bayes' theorem, the posterior probability of Eq.~\eqref{eq:posterior_prob} can be rearranged into the recursive form
\begin{equation}\label{eq:bayes_recursive}
\begin{split}
    &\hat{p}(s_t) \propto 
    \\
    &p(I_{1, t}I_{2, t}|s_t\mathcal{M}^1_{t-1}\mathcal{M}^2_{t-1})\sum^7_{i = 0}p(s_t|s_{t-1} = i)\hat{p}(s_{t-1} = i),
\end{split}
\end{equation}
where we assume that the occurrence of an error is independent of any previous measurements and that $I_{k, t}$ depends on the error state at time $t$ along with past signal values due to auto-correlations. 

This recursive expression describes a Bayesian filter which takes prior information about the error state of the system and updates it based on the transition probabilities $p(s_t|s_{t-1})$ and measurement likelihoods $p(I_{1, t}I_{2, t}|s_t\mathcal{M}^1_{t-1}\mathcal{M}^2_{t-1})$. The filter can be easily implemented once we have functional forms for these two terms, which we describe next.

\subsubsection{Markovian State Transitions}
The Markovian assumption inherent in $p(s_t|s_{t-1})$ is reasonable, given that the net effect of an additional bit-flip error depends only on the error state the system before the error. We assume hereafter that the error rate $\gamma_q$ is identical for all three qubits, i.e., $\gamma_q=\gamma$. This allows us to model the errors as a Markov chain~\cite{Norris_1997} with an $8\times8$ rate matrix $Q$ given by
\begin{equation}\label{eq:Q_bitflip}
    Q_{ij} = 
    \begin{cases}
    -3\gamma & \text{if } j = i \\
    \ \ \gamma & \text{if } j \oplus i \in \{1, 2, 4\} \\
    \ \ 0 & \text{otherwise},
    \end{cases}
\end{equation}
where we define our basis such that index $i\in\{0,\ldots,7\}$ corresponds to the error state whose classical binary representation is equal to $i$, e.g. $5 \rightarrow \ket{101}$. 

Since $Q$ only gives the rate of transition per unit time, we need to compute the transition matrix $J$ in order to get probabilities for a finite step. This matrix can be derived from $Q$ as
\begin{equation*}
    J = e^{Q \Delta t},
\end{equation*}
where $\Delta t$ is the length of the time step. Element $J_{ij}$ gives the probability of transitioning from error state $i$ to error state $j$ across the time step, so we can relate $p(s_t|s_{t-1})$ to $J$ as $p(s_t = j|s_{t-1} = i) = J_{ij}$. Using $J$, the sum in Eq.~\eqref{eq:bayes_recursive} can be evaluated to give probabilities $\tilde{p}(s_t)$
\begin{equation}\label{eq:bayes_markov}
    \tilde{p}(s_t = j) \equiv \sum^7_{i = 0}\hat{p}(s_{t-1} = i)J_{ij},
\end{equation}
which take into account the transitions induced by bit-flip errors during the time step. 

\subsubsection{Measurement Likelihoods}

The measurement likelihood $p(I_{1, t}I_{2, t}|s_t\mathcal{M}^1_{t-1}\mathcal{M}^2_{t-1})$ describes the probability of generating signal values $I_{1, t}$ and $I_{2, t}$ given that the system is in error state $s_t$ and that we had previously measured the values in $\mathcal{M}^1_{t-1}$ and $\mathcal{M}^2_{t-1}$. Since the noise from each syndrome is independent, we can factor the likelihood as
\begin{equation*}
    p(I_{1, t}I_{2, t}|s_t\mathcal{M}^1_{t-1}\mathcal{M}^2_{t-1}) = p(I_{1, t}|s_t\mathcal{M}^1_{t-1})p(I_{2, t}|s_t\mathcal{M}^2_{t-1})
\end{equation*}
with $I_{1, t}$ and $I_{2, t}$ contributing independently to the probability.

If the noise source is assumed to be Gaussian, then the probability density for each $I_{k, t}$ has the form
\begin{equation*}
  p(I_{k, t}|s_t\mathcal{M}^k_{t-1}) =
  \frac{1}{2\pi\sigma^2}\exp\left[
  \frac{-(I_{k, t} - \mu_{k, t})^2}{2\sigma^2}
  \right],
\end{equation*}
where $\mu_{k, t}$ and $\sigma^2$ are the mean and variance of the signal conditioned on the past measurements $\mathcal{M}^k_{t-1}$. In practice the auto-correlations rapidly decay, so we only need to condition on a small number of recent measurements. Hence, we let $m_{k, t-1}$ be the vector of these measurements, and let $c$ be the vector of their corresponding covariance values. Then
\begin{align}
    &\mu_{k, t} = \bar{S}_{k, t} + c^T\Sigma^{-1}(m_{k, t-1} - \bar{S}_{k, t}\vec{1}),\label{eq:corr_mean}
    \\
    &\sigma^2 = \frac{\tau}{\Delta t}  - c^T\Sigma^{-1}c,\label{eq:corr_variance}
\end{align}
where $\vec{1}$ is a vector of ones with the same dimension as $m_{k, t-1}$, $\Sigma$ is the covariance matrix of the variables in $m_{k, t-1}$, and $\bar{S}_{k, t}$ is the mean corresponding to error state $s_t$~\cite{Rue_Held_Held_2005}. Since the system always begins in the coding subspace, each error state maps to a definite error subspace and therefore has definite syndrome values regardless of how the logical state was initialized.

After the measurement pair $I_{k, t}$ is received, the Gaussian likelihood functions are used to convert the probabilities from Eq.~\eqref{eq:bayes_markov} into the next posteriors $\hat{p}(s_t)$ as
\begin{equation}\label{eq:likelihood+markov}
    \hat{p}(s_t) \propto \tilde{p}(s_t)\cdot p(I_{1, t}|s_t\mathcal{M}^1_{t-1}) p(I_{2, t}|s_t\mathcal{M}^2_{t-1}),
\end{equation}
which will become probabilities after normalization.

\subsubsection{Procedure for Error Correction}

The probabilities from Eq.~\eqref{eq:likelihood+markov} can be understood as describing how likely it is that the system is in each of the eight error states based on the judgment of the model. Whenever $\ket{000}$ does not have the highest probability, we can infer that at least one error has occurred and take the appropriate action to correct it. This procedure, which effectively takes the \textit{argmax} of the posteriors, can be altered if certain forms of misclassification are more costly than others, or if the act of making a correction itself carries some cost. The procedure can also be modified so that it is more robust to imperfections in the signal, as we do in Sec.~\ref{sec:rnn} by introducing the $\tau_{ignore}$ and $\tau_{streak}$ hyperparamters.

Whenever any correction is made, we must update the model with this information by permuting its probabilities to reflect the applied bit-flip. In our example, a correction on the second qubit would lead us to swap the probabilities between pairs of error states which differ in only the second qubit, e.g., $\ket{010} \rightleftharpoons \ket{000}$. Without this update the model will continue to recommend the same correction repeatedly, as it does not realize that the state of system has been changed.

A connection can be made between the Bayesian algorithm described here and the maximum likelihood decoder (MLD) commonly used in discrete error correction \cite{Bravyi_Suchara_Vargo_2014}. Given a specific noise channel and qubit encoding, the MLD is the protocol with the greatest probability of successfully correcting an error, assuming that we have access to \textit{projective} measurements of the syndromes. The Bayesian model can be viewed as an extension of the MLD to the continuous measurement regime, where the syndrome measurements provide us with incomplete knowledge of the error subspace. As the variance of the Gaussian measurement noise goes to zero, the Bayesian model reduces to the standard MLD protocol for the three-qubit bit-flip code.

\subsubsection{Impact of Signal Imperfections}
Compared to thresholding schemes, the Bayesian classifier described here is far more sensitive to the assumptions we make about the noise and error distributions. Such sensitivity can be an advantage, since it allows for near optimal performance when our knowledge of these distributions is accurate.

Of course, when our assumptions about the distributions are wrong, the accuracy of the model can suffer significantly. Out of the three imperfections described in Sec.~\ref{sec:setup}, only the auto-correlation of neighboring samples is directly accounted for in the model. The resonator transients occur over relatively short time intervals, so they are likely to have only a modest impact on the model's performance. The syndrome drift also has a negative impact, as the mean values of the Gaussian distributions are key parameters in the model. If there is a discrepancy between the actual signal means and our pre-programmed values, then every measurement likelihood calculation will be biased. 

We explore the size and significance of these effects for all three of our models in Sec.~\ref{sec:experiments}. 

\subsection{Recurrent Neural Network (RNN)}\label{sec:rnn}
Neural networks are a subset of the broader family of machine learning methods based on acquiring a learned representation of the data, which consists of parameterized layers of linear transformations and nonlinear activation functions. RNNs are a class of neural network in which the layers connect temporally, combining the previous time step and a hidden representation into the representation for the current time step. They are thus well suited for representation of the time-dependence of continuously measured error syndromes over discrete time steps. Using a training set of labeled signals, the RNN can learn the properties of the weak measurement signal and the structure of the underlying bit-flip channel, which allows it to accurately detect errors as they occur.

The dynamics of a simple recurrent neural network can be expressed by the following equations:
\begin{equation*}
\begin{split}
&h_{t}=\sigma_{h}\left(W_{h} x_{t}+U_{h} h_{t-1}+b_{h}\right), \\
&y_{t}=\sigma_{y}\left(W_{y} h_{t}+b_{y}\right).
\end{split}
\end{equation*}
For each time step $t$, the network accepts the input vector $x_t$ and, along with the hidden state vector from the previous time step $h_{t-1}$, performs a linear transformation parameterized by the weight matrices $W_{h}$ and $U_{h}$ and the bias vector $b_{h}$ before applying a nonlinear activation function given by $\sigma_h$. The result is the hidden state vector for the current time step $h_{t}$, which is acted upon by an analogous series of operations defined by $W_{y}$, $b_y$ and $\sigma_y$ to produce the output vector $y_t$. We note that the hidden state $h_t$ effectively encodes a description of the history of inputs $\{x_{t'}\}_{t'=0}^{t'=t}$, which therefore allows the network to extract temporal, non-Markovian features from the data.

In our context, we consider the input at each time step to be the vector of measurement signals plus the initial basis state,
\begin{equation} \label{eq:rnn_input}
    x_t = \left[\begin{array}{c}
I_{1,t} \\
I_{2,t} \\
s_0 \\
\end{array}\right].
\end{equation}
Moreover, instead of the standard recurrent neural network architecture, we use a long short-term memory network (LSTM)~\cite{hochreiter_lstm_1997}, which is a particular type of recurrent neural network that involves cell states and various gates to evade the vanishing gradient problem of standard RNN architecture~\cite{kolen_gradient_2001}. Nevertheless, the same principle underlying the standard function of RNN applies. The output $y_t$ of the LSTM layer is subsequently passed through a dense layer and a softmax activation to produce the posterior probabilities of the eight basis states $p(s_t | \mathcal{M}^k_t)$, and we select the basis state with the highest posterior as the prediction $\hat{s}_t$.

\subsubsection{Training}
Training samples for the RNN require accurate labeling of the states corresponding to the measurement signals at every time step. However, in reality, decoherence effects such as amplitude damping and thermal excitation prevent us from knowing the correct state of the system at some arbitrary time. As a result, to train the RNN, we have to resort to measurement signals with a well defined underlying quantum state. This can be achieved by simulating the measurement signals on states in the absence of unwanted decoherence effects, which will be described in details in Sec~\ref{sec:experiments}. In the simulations, we provide the measurement strength, the single-qubit bit-flip error rate and the initial quantum state as input parameters, and the simulation produces a large number of quantum trajectories to be the training samples of the RNN. We then train the RNN to diagnose bit-flip errors on the three-qubit system, and the trained RNN can be subsequently used to actively correct for errors that occurred. That said, the same information used to generate the training samples is also provided as prior knowledge to the double threshold and the Bayesian model. The two models both require an explicit estimation of the measurement strength as well as the assumption of a certain error rate.

We maximize the likelihood of the RNN parameters on the training set by minimizing the cross entropy batch total loss function, which is defined as
\begin{equation} \label{eq:loss_function}
    \mathcal{L} = -\frac{1}{NT} \sum_{n=1}^N \sum_{t=1}^{T} \log{p_{n}(s_t)},
\end{equation}
where $p_{n}(s_t)$ stands for the posterior probability of the true basis state $s_t$ at time step $t$ in the $n$-th sample, while $N$ denotes the mini-batch size and $T$ denotes the total number of steps in each training sample. 

To update the parameters to minimize the loss, we perform an iterative training procedure where for each step and parameter $w$, one applies a gradient descent update of the form $w \leftarrow w - \eta(\partial \mathcal{L}/\partial w)$,
where the gradients ${\partial \mathcal{L}}/{\partial w}$ are computed via backpropagation through the computation graph of the network. 

In our experiments, the gradient descent update is preformed using the ADAM optimizer~\cite{Kingma2015AdamAM}. We adopt a two-layer stacked LSTM with a hidden state size of $32$. This small hidden size limits the largest matrix-vector multiplication in computations, hence the memory required, and also limits the number of parameters, facilitating the implementation of the network in real-time experiments. We further provide a comparison test on the performance of different hidden state sizes in App.~\ref{app:rnn_hidden_performance} and show that both smaller LSTM and gated recurrent unit (GRU) architecture~\cite{gru_2014} offer comparable performance for our purpose. The number of stacked layers of the LSTM/GRU and the hyperparameters, such as the batch size in training, are tuned with the assistance of Ray Tune~\cite{liaw2018tune}. 

\subsubsection{Re-calibration Method for Error Correction}
When performing active error correction, we once again wish to avoid the delay in the posterior probabilities output by the network to reflect the application of an error correction operation $C_{\text{op}}$ on the system. In the case of the Bayesian classifier, we permute the elements of the vector of posterior probabilities, which encodes the state of the model, in accordance with the error correction operation. For the RNN, however, we cannot apply a particular transformation to the hidden state such that the vector of posterior probabilities outputted by the network is permuted in analogous manner, since the function mapping the hidden state to the output vector of posterior probabilities is highly nontrivial.

Any such delay in the network remaining unaware of the quantum state having been corrected is harmful, because another error $X_q$ occurring during this delay, compounding with the correction $C_{\text{op}}$ on the first error, will induce a logical error at the next error correction operation. To see this clearly, considering that the physical qubits are initially in $\ket{000}$, and the first error $X_1$ results in the state $\ket{100}$. After detecting the error, the model makes a correction that instantly returns the state back to $\ket{000}$. However, the RNN still has the knowledge of the qubits being in $\ket{100}$ until some time later at $t_{\text{realize}}$ before accepting sufficient number of $x_t$'s that allows it to predict $\ket{000}$. If a second error $X_2$ occurs before $t_{\text{realize}}$, the syndromes become $(S_1=-1, S_2=-1)$ because the state becomes $\ket{010}$, whereas the RNN, only knowing the state in $\ket{100}$, will eventually predict $\ket{101}$ that has the same syndromes, which is then equivalent to diagnosing an $X_3$ error. After applying a second error correction $C_{\text{op}}=X_3$, the physical qubits are now in $\ket{111}$, constituting a logical error. In other words, since we are not capable of injecting the knowledge of a correction operation into the RNN, a correction operation is equivalent to an error seen by the RNN and active correction effectively increases the bit-flip error rate $\gamma$ in the eyes of the network. Although the correction is correlated with the detected error, the network is generally trained on quantum trajectories with uncorrelated random bit-flip error instances. As will be explained in \ref{sec:state_tracking} that a greater $\gamma$ will induce more logical errors, we conclude that the naive approach of active correction with the RNN suffers from more logical errors.  

Therefore, we propose the following re-calibration protocol to effectively hide the action of any error correction operation from the network, so that there is no longer any delay in the posterior probabilities to begin with.

We specifically keep track of all the error correction operations that has been applied up to the present $t$, 
\begin{equation*}
N_{q, t} = \text{Number of } {X_q} \text{ corrections applied}.
\end{equation*}
When the measurement signals $I_{1,t}$ and $I_{2,t}$ have symmetric noise around their respective mean values and the possible means of $I_{k, t}$ are always equal and opposite, each $C_{\text{op}}$ correction changes the mean of $I_{1, t}$ by a factor of $-1$ if $C_{\text{op}}=X_1$, changes the mean of $I_{2, t}$ by a factor of $-1$ if $C_{\text{op}}=X_3$, and changes the mean of both $I_{k, t}$ by a factor of $-1$ if $C_{\text{op}}=X_2$. To hide all the corrections done in the past, the measurement signals that are provided as input to the network for all subsequent time steps are then flipped according to $N_{q,t}$,
\begin{equation*}
\begin{split}
    I'_{1,t} = (-1)^{N_{1, t} + N_{2, t}} I_{1,t},\\
    I'_{2,t} = (-1)^{N_{2,t} + N_{3,t}} I_{2,t},
\end{split}
\end{equation*}
which we called the re-calibrated signals. From the perspective of the RNN when taking in $I'_{k, t}$, it appears as if no error correction operation has been applied to the physical qubits. 

Given that at some time step we predict a different state $\hat{s}_t$, we now perform our error correction operation relative to the previous predicted state $\hat{s}_{t-1}$.

\subsubsection{Adaption to Resonator Transients for Probabilistic Models}
When the possible means of $I_{k, t}$ are not equal and opposite, as occurs in the resonator transients upon applying $C_{\text{op}}$, the re-calibration method breaks down, because flipping the means of either or both $I_{k, t}$ does not produce the means as if there was no correction applied. A solution to this is to impose an ignore time period $\tau_{\text{ignore}}$ right after the correction is applied at some $t$. During $(t, t+\tau_{\text{ignore}}]$, no input $x_t$ is fed into the RNN. As a result, the hidden state of the network is frozen until the ignore time period ends. The re-calibrated signals are accepted by the network only after $t+\tau_{\text{ignore}}$, which reduces the risk of getting incorrect predictions during the transients, but effectively increases the detection time of any error that occurs during the ignore period.

Imposing $\tau_\text{ignore}$ should be accompanied by a measure to ensure that the RNN diagnoses any error with sufficiently high confidence so that fewer false alarms of error will be followed by an ignore period $\tau_\text{ignore}$ upon correction. A feasible measure in practice is to determine an error correction operation only if the RNN predicts the same state $\{\hat{s}_{t'}\}_{t'=t}^{t'=t+\tau_\text{streak}}$ for a streak of time steps $\tau_\text{streak}$ that is different from the old state $\hat{s}_{t-1}$, which is a discrete quantity that is easy to optimize. The $\{\tau_\text{ignore}, \tau_\text{streak}\}$ then constitutes a minimal set of tunable hyperparameters for the task of active correction in the presence of resonator transients, which applies to the Bayesian classifier explained in Sec.~\ref{sec:bayesian} as well.

\section{Simulated Experiments}\label{sec:experiments}

To evaluate the effectiveness of the three models described in Sec.~\ref{sec:models}, we test their error correction capabilities on a large number of synthetic measurement sequences. The motivation for using artificial data instead of real data is twofold. First, by using artificial data we can precisely control the underlying measurement distribution, which allows us to separate out the effects of the different imperfections identified in Sec.~\ref{sec:setup}. Second, it is important that we know the true state of the system at every time step, as this is necessary both to train the RNN and to calculate intermediate fidelity values. Such knowledge would not be possible on a near-term quantum computer due to strong undesirable decoherence.

To ensure that our simulations are grounded in reality, we model them on data taken from a superconducting qubit device. Fig.~\ref{fig:signal_example} shows measurements taken from this reference data, which consists of approximately $1.6\times 10^6$ sequences lasting \SI{6}{\micro\second} each \footnote{The $1.6\times 10^6$ sequences break down to about $50,000$ sequences for each of the eight initial states and for each of the $X_1$, $X_2$, $X_3$ injected bit-flip or no injected bit-flip.}. The sequences are comprised of $192$ measurement pairs (one for each resonator), sampled every \SI{32}{\nano\second}. The data contains both ``flat'' sequences, in which no bit-flip occurs, as well as sequences in which a bit-flip is deliberately applied to one of the three qubits to induce a state transition. Since these bit-flips are all applied at precisely the same time, we are able to track how the the signal mean changes during the transient period.

Across all of our tests we employ four different simulation schemes, each of which is described below. The schemes are designated with letters A--D in order of how much non-ideal behavior they include, with Scheme A having no imperfections and Scheme D having all three imperfections. In all schemes, we ignore the thermal excitation for each qubit, since a typical excitation rate is on the order of \SI{1}{\per\milli\second}.

\subsubsection{Scheme A: Idealized Behavior}
In our first scheme, the simulated signal simply conforms to the idealized behavior given by Eq.~\eqref{eq:signal_expression}. At the beginning of each measurement sequence the system is set to a specified initial state in the coding subspace, and then the state of the next time step is determined by sampling a number $n_q$ of bit-flips $X_q$ for each qubit from the Poisson distribution, such that $n_q = \exp(-\gamma\Delta t) (\gamma\Delta t)^{n_q}/n_q!$ where $\Delta t$ is the time step size. These errors are applied to the corresponding qubits to get the next state. This cycle of sampling and propagating errors is repeated until we have generated a sufficiently long sequence of states. 

To create the corresponding $I_{k, t}$, we sample a uni-variate Gaussian distribution at each time step with variance $(\Gamma^k_m \Delta t)^{-1}$ and a mean of $\pm 1$ determined by the syndrome eigenvalue at that step. Our reference data has
\begin{equation*}
    \Gamma_m^k\approx4.7\times10^6\ \SI{}{\per\second},\quad
    \Delta t = 32\times 10^{-9}\ \text{s},\quad
    \eta_k \approx 0.5, 
\end{equation*}
where $\Gamma_m^k$ needed to be estimated from the measurement signals while $\Delta t$ was known to us in advance. This sequence of Gaussian samples plus the underlying states provides a complete description of a system in the context of our error correction task.

\subsubsection{Scheme B: Auto-correlations}

As a first step away from ideal behavior, we consider noise that is correlated across time. The data generation process for this scheme is effectively the same as that of Scheme A, except that the noise must be sampled sequentially in order to correctly capture the auto-correlations. In our reference data we find that significant auto-correlations extend back roughly four steps, with covariance given by
\begin{equation*}
    c^T_k \approx 5.94 \cdot
    \begin{bmatrix}
         0.61 & 0.25 & 0.1 & 0.05
    \end{bmatrix}
\end{equation*}
whose $i$th element is at lag-$i$. These values were found by taking every contiguous subsequence of length five in our reference data and using them all to compute a covariance matrix. We can simulate Gaussian noise with these auto-correlations one step at a time using Eqs.~(\ref{eq:corr_mean}, \ref{eq:corr_variance}).

\subsubsection{Scheme C: Auto-correlations with Resonator Transients}

For our third scheme, we keep the auto-correlations from Scheme B but alter the behavior of the syndrome values so that they include the resonator transients seen in Fig.~\ref{fig:signal_example} and explained in App.~\ref{app:transients}. To incorporate these patterns into our simulation, we first extract the mean values of the transient patterns from our reference data, consisting of $94$ steps in total, for each of the twenty-four different single-flip transitions. Our sequence generation process is then identical to Scheme B, except that after an error occurs the next $94$ measurements are sampled from Gaussians centered on the transient means instead of the syndrome eigenvalues. The pattern that we use is matched to the state of the system before and after the error. After the transient period has elapsed, the means are set back to $\pm 1$ and further samples are generated as usual until another error occurs.

\subsubsection{Scheme D: All Imperfections}
Our final simulation scheme takes the auto-correlations and resonator transients from Scheme C and adds an underlying drift term to the the syndrome means. Since our reference data contains over a million trajectories collected over the span of multiple hours, it is possible to observe significant differences in the syndrome means between trajectories that are separated by large amounts of time, possibly due to temperature fluctuations. 

For our experiments we elected to apply a linear drift $\Delta_i$ governed by 
\begin{equation*}
    \Delta_i = \frac{0.4}{N} \cdot i,
\end{equation*}
where $i$ is an index that arbitrarily orders the different measurement sequences that we generate and $N$ is the total number of these sequences. This drift term is added to every measurement in the $i$th sequence, resulting in a uniform shift of the overall signal means. The net drift across all runs represents a $40\%$ change, which is consistent with the magnitude of the drift observed in our reference data.

\subsection{Quantum Memory State Tracking}\label{sec:state_tracking}
\begin{figure*}
 \centering
\includegraphics[scale=0.47]{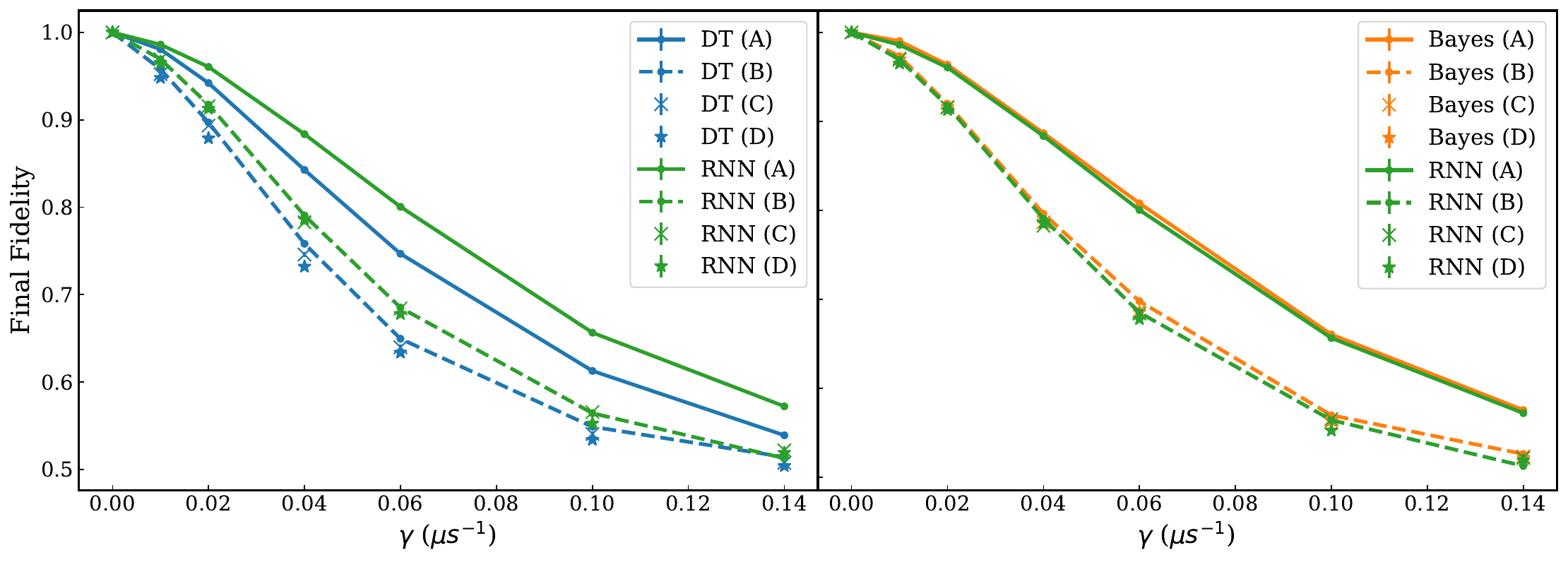}
 \caption{The final fidelity with respect to the initial state $\ket{000}$ in Schemes A, B, C, D with the double threshold (DT), Bayesian and RNN classifier, as a function of single-qubit bit-flip rate $\gamma$ at an operation time $T=\SI{20}{\micro\second}$. Each data point is averaged over $30,000$ quantum trajectories. For better visualization, we split the figure into two plots, with the left one comparing the RNN classifier to the double threshold, and the right one comparing the RNN classifier to the Bayesian classifier. On the left, we see that the RNN classifier outperforms the double threshold in all schemes. Whereas on the right, it shows that the RNN approximates the Bayesian classifier, which is the optimal one among the three, in all schemes. The error bars show the standard error of the mean.}
 \label{fig:state_tracking}
\end{figure*}

In quantum memory, it suffices to track the basis states in response to the bit-flip errors that have occurred and only apply error correction operations when needed. We generated $30,000$ trajectories of length $T=\SI{20}{\micro\second}$ from all four simulation schemes with a pre-defined single-qubit error rate as our testing samples, among which are equal portions of trajectories initialized in one of the eight basis states. While the RNN model employed here is trained on $100,000$ quantum trajectories from the corresponding simulation scheme, the error rate, noise variance and auto-correlations input to the Bayesian model are also estimated from those quantum trajectories. The tunable parameters in the double threshold model are numerically optimized in schemes with imperfections; the filtering time $\tau$ typically lies in the range  $\SI{0.3}{}-\SI{1.6}{\micro\second}$, with larger $\tau$ for smaller $\gamma$. 

In Fig.~\ref{fig:state_tracking}, we compare the final fidelity $\mathcal{F}=\abs{\braket{\psi_T|\psi_0}}^2$ against the initial state of the three models in tracking these quantum trajectories subject to bit-flips. The trend is that the final fidelity decreases as a function of the single-qubit error rate $\gamma$. This is because the higher the error rate is, the more chances there will be two different bit-flips before the correction to the first bit-flip is made, resulting in a logical error upon the correction, and therefore a lower final fidelity. For instance, a state starting at $\ket{000}$ is flipped to $\ket{001}$ at $t_1$ and is later also flipped to $\ket{011}$ at $t_2>t_1$, such that $t_2$ is smaller than $t_1+t_\text{detect}$ where $t_\text{detect}$ is the detection time of the first error. Subsequently, the model perceiving syndromes with $(S_1=-1,S_2=+1)$ will eventually make a $C_\text{op}=X_1$ correction and change the state to $\ket{111}$, leading to a logical error. From the above argument, it is also evident that a shorter detection time is beneficial. 

From Fig.~\ref{fig:state_tracking}, we see that the RNN and the Bayesian classifier outperform the double threshold in all simulation schemes, whereas the RNN approximates the Bayesian classifier in all schemes. As discussed in Sec.~\ref{sec:bayesian}, the Bayesian classifier is the optimal model of the three in Schemes A and B where there are only auto-correlations in the signals, which is validated in this task. The fact that their performances in Schemes C and D are very similar to that in Scheme B indicates that the resonator transient pattern and the drifting of the means do not have a significant effect on all three models. 

It is reasonable that the drift has a small negative effect to the two probabilistic models, since the drift is usually on the order of the separation of mean values of the two parities, which is in turn one order of magnitude smaller than the standard deviation of the noise. The large noise variance obscures the drifting means, making the drifted signals appear like more noisy signals with fixed means.

\subsection{Extending \texorpdfstring{$T_1$}{Lg} Time of the Logical Qubit}

\begin{figure*}
 \centering
  \includegraphics[scale=0.43]{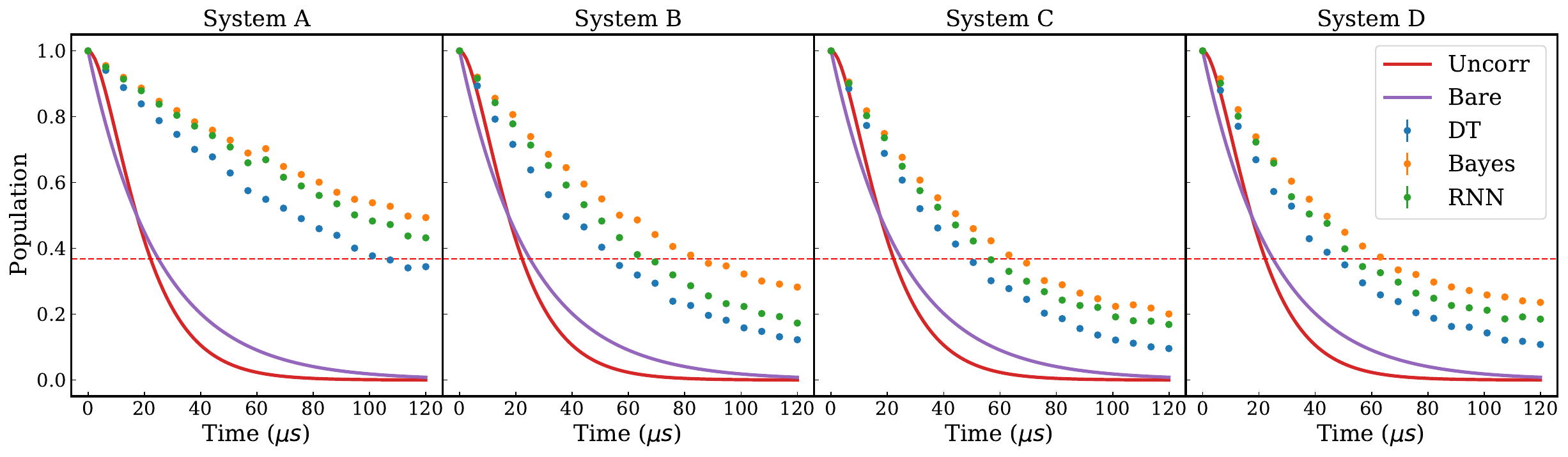}
 \caption{The population of the excited states $\{\ket{111}, \ket{110}, \ket{101}, \ket{011}\}$ as a function of time, obtained from simulated experiments with the four different schemes at a single-qubit decay rate of $\gamma=\SI{0.04}{\per\micro\second}$. Each data point is averaged over $3,000$ independent quantum trajectories. The three-qubit system is initialized to $\ket{1}_\text{L}=\ket{111}$. As a comparison, the bare qubit (purple curve) is initialized to the $\ket{1}$ state and is subject to amplitude damping with a time constant of $T_1=\SI{25}{\micro\second}$, i.e., a decay rate of $\SI{0.04}{\per\micro\second}$. For reference, the uncorrected three-qubit system decay curve is shown in red (see App.~\ref{app:population_curve}). For all schemes, the RNN-based model outperforms the double threshold model.}
 \label{fig:t1_decay}
\end{figure*}

Although the models are motivated by correcting bit-flip errors, they can also be exploited in extending the $T_1$ time of the logical qubit in $\ket{1}_\text{L}=\ket{111}$. For this task, actively correcting the state is required as opposed to merely tracking the state. While for practical purpose the RNN model is trained on $30,000$ quantum trajectories under bit-flips with a length of $T=\SI{120}{\micro\second}$, the Bayesian model, whose parameters are estimated from the same set of trajectories, uses a different transition matrix generated by $Q'$ shown in Eq.~\eqref{eq:t1_Q_transition} which takes into account the asymmetric probabilities of transitions between the ground and excited state. The parameters for the double threshold model is numerically optimized on the same set of quantum trajectories.

For the three-qubit system initialized to the fully excited state $\ket{111}$, we inspect the population within a Hamming distance $1$ away from the initial state, i.e., the population $P_\text{exc}$ of the set of basis states $\{\ket{111}, \ket{110}, \ket{101}, \ket{011}\}$, since these states can be recovered to the initial state by a majority vote. We compare this $P_\text{exc}$ against the population of the excited state $\ket{1}$ of a bare qubit as a function of time in all four simulation schemes, and the results are shown in Fig.~\ref{fig:t1_decay}. In all schemes, the encoded three-qubit system $P_\text{exc}$ decays much slower under active correction by any of the three models than the bare qubit excited state population. In all schemes, both the Bayesian and the RNN-based model outrun the double threshold model. 

\subsection{Protecting against Bit-flip Errors}
\begin{figure*}
 \centering
\includegraphics[scale=0.43]{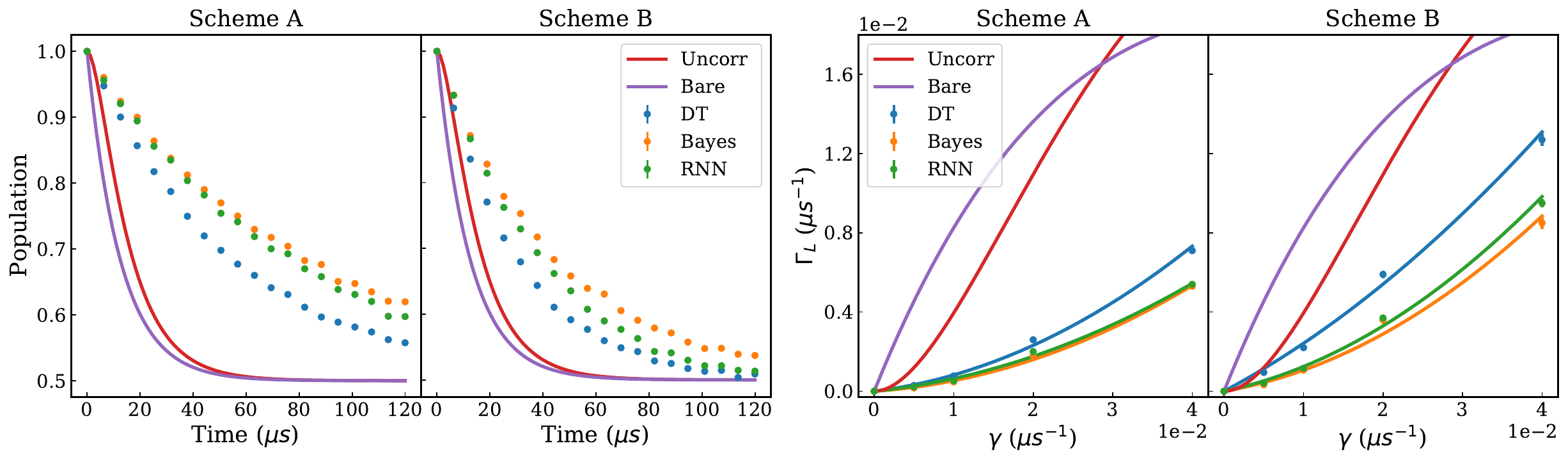}
 \caption{Left: the population of the excited states $\{\ket{111}, \ket{110}, \ket{101}, \ket{011}\}$ as a function of time, obtained from simulated experiments under Schemes A and B at a single-qubit bit-flip rate of $\SI{0.04}{\per\micro\second}$. Each data point is averaged over $3,000$ independent quantum trajectories. The three-qubit system is initialized to $\ket{1}_\text{L}=\ket{111}$. As a comparison, the bare qubit (purple curve) is initialized to $\ket{1}$ and is subject to a bit-flip rate of $\gamma=\SI{0.04}{\per\micro\second}$. As a reference, the uncorrected three-qubit system decay curve is shown in red (see App.~\ref{app:population_curve}). In Schemes A and B, the Bayesian model is the best among the three, and the Bayesian and RNN-based model both outrun the double threshold model. Right: the initial logical error rate $\Gamma_\text{L}$ at $\SI{9.6}{\micro\second}$ as a function of the single-qubit error rate $\gamma$. The fitted quadratic curves show a strong suppression of $\Gamma_\text{L}$ for all three models in both schemes.}
 \label{fig:bitflip}
\end{figure*}

Similar to the task of extending the $T_1$ time of the state $\ket{1}_\text{L}$, here we employ the three models to protect the initial state $\ket{1}_\text{L}$ from bit-flips. Shown in Fig.~\ref{fig:bitflip}, we compare the population $P_\text{exc}$ of the three-qubit system against the excited population of the bare qubit in time. For Schemes A and B, both the Bayesian and the RNN-based model have an advantage over the double threshold. Furthermore, in Fig.~\ref{fig:bitflip} we extract the initial logical error rate $\Gamma_\text{L}$ as a function of $\gamma$ by computing the time derivative of $P_\text{exc}$ at $\SI{9.6}{\micro\second}$ at each $\gamma$. In either scheme with any of the three models, $\Gamma_\text{L}$ scales approximately quadratic in $\gamma$, and we can see a strong suppression of $\Gamma_\text{L}$ relative to a bare qubit or the uncorrected three qubits. We remark that, by introducing feedback based on noisy weak measurements, any correction protocol can underperform a majority vote on the encoded qubits without error correction at sufficiently small $\gamma$ or runtime.

\begin{figure}
    \centering
    \includegraphics[width=0.38\textwidth]{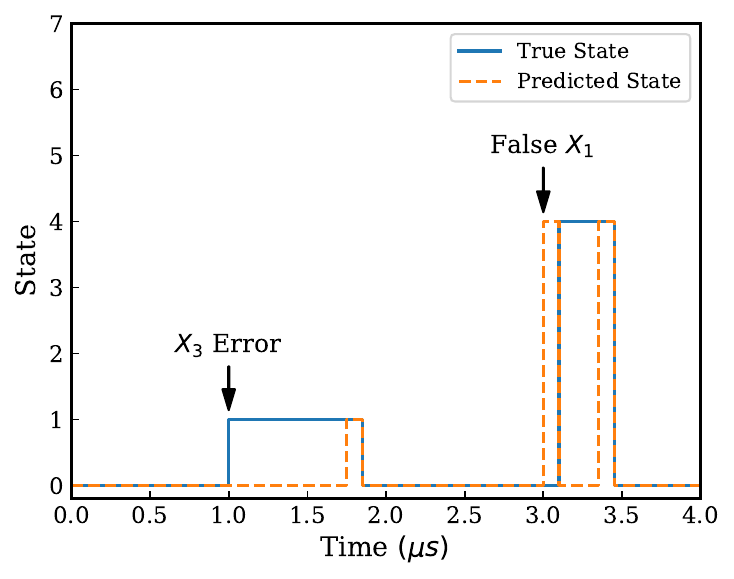}
    \caption{Response of the system basis state and model to a true bit-flip error and a false alarm as a function of time. At \SI{1.0}{\micro\second} an $X_3$ error is applied to the system, and after a small delay the error is detected and corrected. At \SI{3.0}{\micro\second} the model falsely detects and then ``corrects'' for an $X_1$ error, which results in the system being temporarily pushed into an error subspace before the mistake is recognized and corrected. There are visible small constant offsets between the prediction and the system state at the false alarm due to the streak time period imposed in the correction protocol.}
    \label{fig:error_response}
\end{figure}

To better understand the performance of the models in this important task, we analyze the detection time spent in true positive detection as well as the number of false alarms when the three-qubit system is in $\ket{1}_\text{L}$. The difference between a true positive and a false alarm is illustrated in  Fig.~\ref{fig:error_response}, which shows the actual and predicted states of the system when an $X_3$ error occurs and when the model falsely detects an $X_1$ error. When a true error occurs, the system remains in the corresponding error subspace for a duration determined by the detection time of the model, after which the error is corrected. By contrast, when the model falsely detects that an error has occurred due to measurement noise, it improperly applies a bit-flip to the system and thus pushes it out of the code subspace. After more measurements are recorded, the model determines that the system is in an error subspace and fixes its mistake by applying another bit-flip.

\begin{figure}
 \centering
 \includegraphics[scale=0.62]{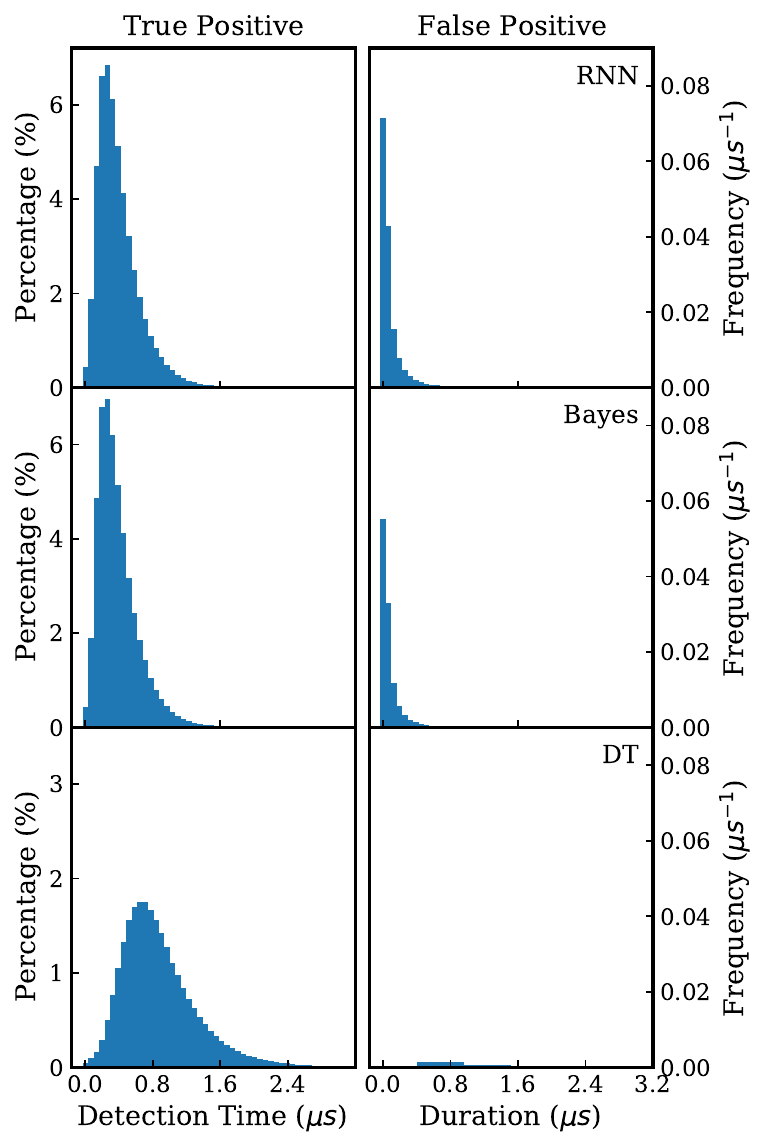}  \caption{The distribution of detection time (with the left $y$-axis) and the distribution of false alarms of bit-flips (with the right $y$-axis) when the state is originally in $\ket{111}$, over $100,000$ quantum trajectories with an operation time $T=\SI{120}{\micro\second}$ and with a single-qubit bit-flip rate $\gamma=\SI{0.04}{\per\micro\second}$. The three qubits are initialized to $\ket{111}$. The overall frequencies of all false alarms for the RNN-based, Bayesian, and double threshold models are $0.155(5)$, $0.117(2)$, $\SI{0.0022(2)}{\per\micro\second}$, respectively.}
 
 \label{fig:hist}
\end{figure}

As explained in Sec.~\ref{sec:state_tracking}, a shorter detection is favorable and will lead to better error corrections, whereas here we can expect more frequent false alarms arises for models with a shorter detection time as a trade off, since the model is prone to make a correction. This is demonstrated in Fig.~\ref{fig:hist}, where we can see that the best two models, the Bayesian and the RNN-based, both have a shorter detection time and more frequent false alarms at the same time. Nevertheless, for both of these two models, the overall frequency of all false positive detection remains low and is on the order of \SI{0.1}{\per\micro\second}.

\subsection{Quantum Annealing with Time-dependent Hamiltonians}

Having demonstrated a clear advantage using the RNN-based protocol for tasks in the quantum memory setting over the double threshold protocol, we now study the performance of our protocol for quantum annealing, using a time-dependent Hamiltonian that does not non-commute with the bit-flip errors. We note that the protocol is also applicable to evolution under quantum gate operations. 

In quantum annealing, it is imperative to perform error diagnosis and correction in a manner that is both fast and accurate, in order to avoid accruing these logical errors while single bit-flip errors are being diagnosed and corrected. This is because the action of an error $X_q$ effectively transforms the Hamiltonian from $H(t)$ to $X_q H(t)X_q$ in the Heisenberg picture. Until the error is properly diagnosed and corrected, subsequent coherent evolution of the logical state in the code subspace is due to the modified Hamiltonian $X_q H(t) X_q$. If the original Hamiltonian does not commute with the error, i.e. $X_q H(t)X_q \neq H(t)$, then such evolution will be spurious rather than as originally intended, causing logical errors to accrue.

For this simulated experiment (see App.~\ref{app:annealing}), the annealing Hamiltonian with a strength $\Omega_0$ evolving $\rho_0=\ket{\psi_0}\bra{\psi_0},\ \ket{\psi_0}=(\ket{0}_\text{L}+\ket{1}_\text{L})/\sqrt{2}$ is chosen to be
\begin{equation}\label{eq:annealing_hamiltonian}
    H(t)=-\Omega_0\left[a(t)X_1X_2X_3+b(t)\frac{Z_1+Z_2+Z_3}{3}\right],
\end{equation}
where $a(t)=1-t/T$ and $b(t)=t/T$. In the code subspace, it is equal to
\begin{equation}\label{eq:bare_hamiltonian}
    h(t)=-\Omega_0\left[a(t)\sigma_x+b(t)\sigma_z\right],
\end{equation}
whereas in any error subspace it is equal to the spurious Hamiltonian,
\begin{equation*}
    h_\text{spurious}(t)=-\Omega_0\left[a(t)\sigma_x+b(t)\frac{\sigma_z}{3}\right].
\end{equation*}

We adopt the reduction factor~\cite{atalaya_continuous_2021} as the metric for evaluating the model performance, which is defined as,
\begin{equation}
    \mathcal{R}=\frac{1-\mathcal{F_\text{une}}}{1-\mathcal{F}},
\end{equation}
whose numerator is the final infidelity of an unencoded bare qubit initialized to $\ket{0}$ under the annealing Hamiltonian Eq.~\eqref{eq:bare_hamiltonian}, and whose denominator is the final infidelity of the three-qubit encoded state in the code subspace with respect to the target quantum state. As $\dot{a}(t), \dot{b}(t)\rightarrow 0$, the target quantum state becomes the ground state of the target Hamiltonian.

\begin{figure}
 \centering
 \includegraphics[scale=0.56]{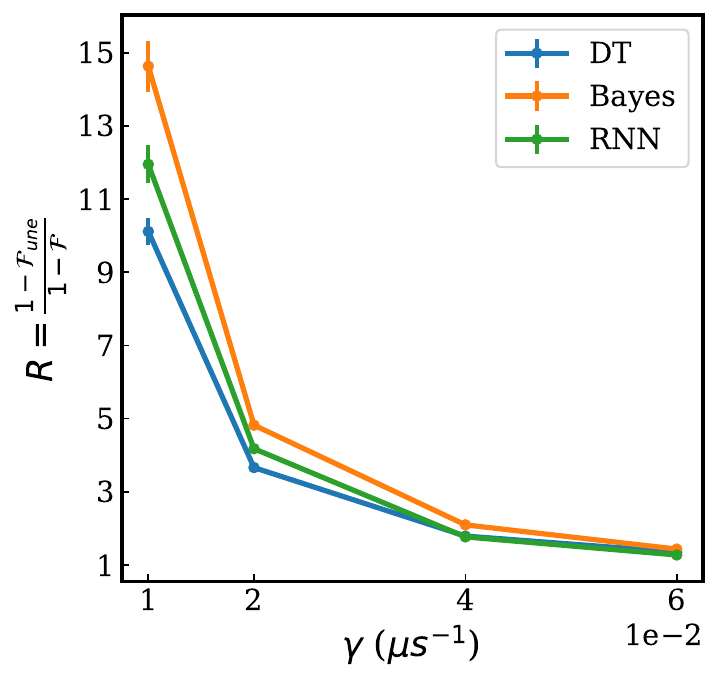}
 \caption{The final infidelity reduction factor as a function of single-qubit bit-flip rate $\gamma$, with an operation time $T=\SI{120}{\micro\second}$, and the strength of the annealing Hamiltonian in Eq.~\eqref{eq:annealing_hamiltonian} equal to $\Omega_0=0.04\Gamma_m$ where the measurement strength is set to $\Gamma_m=\SI{4.7}{\per\micro\second}$. The quantum efficiency is set to $\eta=0.5$. Each data point is averaged over $10,000$ quantum trajectories.
 }
 \label{fig:annealing}
\end{figure}

As shown in Fig.~\ref{fig:annealing}, at relatively low $\gamma$, the Bayesian model achieves the highest reduction factor in Scheme A, while both the Bayesian and the RNN-based model outperform the double threshold. However at sufficiently high error rates $\gamma$, the encoded qubits under active correction using any of the three models show no improvement over a single unencoded qubit, as expected.

\section{Discussion}\label{sec:discussion}

We have proposed an RNN-based CQEC algorithm that is able to outperform the popular double threshold algorithm across all tasks for each of the four simulation schemes tested in Sec.~\ref{sec:experiments}. This result holds regardless of whether the algorithms are protecting a system from bit-flip errors or from amplitude damping, and applies in the case of both quantum memory and quantum annealing. The relative performance of the three models does not depend significantly on the underlying error rate or the duration of the experiment, unless either of these values is exceptionally large.  

The mathematical simplicity of Eq.~\eqref{eq:signal_expression} is a product of many idealized assumptions, so we can expect that measurements taken from real quantum devices will not necessarily be as easy to describe. Our analysis of superconducting qubit measurements in Sec.~\ref{sec:setup} reveals several examples of non-ideal behavior in both the syndrome and noise distributions, and we expect similar findings in the outputs of other devices. While some signal imperfections can be accounted for in traditional CQEC algorithms, such as the incorporation of auto-correlations into the Bayesian classifier, most of them will not be easy to precisely characterize. It is in these situations that neural networks can best demonstrate their advantage, since they do not require any a priori description of the patterns within the measurement signals, but instead learn them directly from the training data. An interesting direction for further study is the extension of the RNN-based CQEC algorithm to correlated and leakage errors.

A CQEC algorithm should be practical to run on a sub-microsecond timescale, typically using an FPGA or other programmable, low-latency device. The Bayesian model requires division to normalize the posteriors, which is a very costly operation on FPGAs. This makes it challenging to efficiently implement the Bayesian model, although a more practical log-Bayesian approach has recently been developed \cite{Convy_Whaley_2021}. The RNN-based model, by contrast, does not require division and avoids this problem. There are many precedents for running RNNs on FPGAs (see e.g.~\cite{Chang_Martini_Culurciello_2016}). Since the RNN architecture used in our paper is small in size (more simplifications are discussed in App.~\ref{app:rnn_hidden_performance}), its computational latency is sub-microsecond. Nevertheless, more work will be needed in order to determine how best to interface the RNN with the quantum computer in a feedback loop. For supervised learning, there is the need for generating a sufficient amount of training data that incorporates the error information and the signal features. Further work could focus on determining the minimum amount and type of data that the RNN needs to train effectively, and understand how these needs change as the number of physical qubits in the error code increases.

Given low-latency implementations of the Bayesian and RNN-based models, an obvious next step for future work would be a direct comparison between these CQEC protocols and existing discrete QEC protocols on quantum hardware. Rist\`e \textit{et al.}~\cite{Riste_2020} have already demonstrated discrete QEC for a three-qubit bit-flip code on transmons, and recent work by Livingston \textit{\textit{et al.}}~\cite{Livingston_2021} has implemented a triple threshold CQEC protocol on similar hardware. By running experiments on a given physical device, a full comparison between discrete and continuous CQEC can be made under realistic conditions. Due to the lack of both entangling gates and ancillas, we are optimistic that CQEC could significantly improve the speed and fidelity of many QEC codes. 

\section*{Acknowledgement}
We would like to thank Philippe Lewalle, John Preskill, Kai-Isaak Ellers, and John Paul Marceaux for helpful discussions.~H. L. and I. C. were supported by the National Aeronautics and Space Administration under Grant/Contract/Agreement No.~80NSSC19K1123 issued through the Aeronautics Research Mission Directorate. S. Z., H. N. N., and K. B. W. were supported by the U.S. Department of Energy, Office of Science, National Quantum Information Science Research Centers, Quantum Systems Accelerator.~W. P. L. and I. S. were supported by the U.S. Army Research Laboratory and the U.S. Army Research Office under Contract/Grant No.~W911NF-17-S-0008. Publication made possible in part by support from the Berkeley Research Impact Initiative sponsored by the UC Berkeley Library.

\bibliography{refer}

\section*{Appendices}
\begin{appendix}

\section{Source Code}
The code developed for all models and simulated experiments can be found \href{https://colab.research.google.com/drive/1NSwz4Qy3SlfE-fptz59-hj3880QJ7DVj?usp=sharing}{here}. Use of the code for any publication should reference this paper. The data that support the findings of this study are available upon request.

\begin{figure*}
  \includegraphics[scale=0.41]{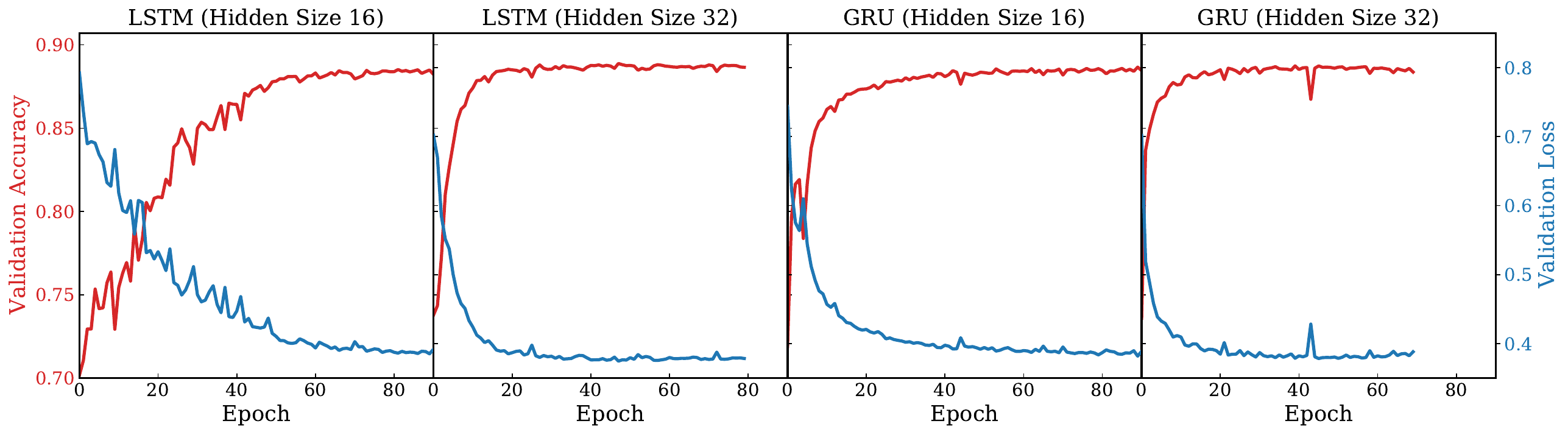}
 \caption{The learning curves of LSTMs with hidden sizes of $16$ and $32$, and of GRUs with hidden sizes of $16$ and $32$, on the state tracking task in quantum memory as described in Sec.~\ref{sec:state_tracking}. The accuracy is defined to be the fidelity with respect to the initial state averaged across all time steps, and the loss is computed by Eq.~\eqref{eq:loss_function}.}
 \label{fig:learning_curves}
\end{figure*}

\begin{table*}[t]
    \caption{The testing performance of LSTM (left) and GRU (right) with different hidden sizes and the corresponding number of trainable parameters. The testing performance is measured by the final excited states population $P_\text{exc}$. The hidden size determines the largest matrix-vector multiplication operation performed when computing the model.}
    \vspace{10pt}
    \begin{minipage}{.5\linewidth}
      \centering
        \begin{tabular}{l|rrrr}
            Hidden size & 8 & 16 & 32 & 64\\
            \hline
            Parameter count & 1064 & 3256 & 13448 & 51464\\
            Final $P_\text{exc}$ ($\pm0.002$) & $0.851$ &  $0.880$ & $0.884$ & $0.882$
        \end{tabular}
    \end{minipage}%
    \begin{minipage}{.5\linewidth}
      \centering
        \begin{tabular}{l|rrrr}
            Hidden size & 8 & 16 & 32 & 64\\
            \hline
            Parameter count & 816 & 2776 & 10152 & 38728\\
            Final $P_\text{exc}$ ($\pm0.002$) & $0.816$ & $0.880$ & $0.879$ & $0.881$
        \end{tabular}
    \end{minipage} 
    \label{tab:rnn_hidden_size}
\end{table*}

\section{Homodyne Measurements} \label{app:homodyne}
The interaction Hamiltonian for the transmission line and the cavity field is given by
\begin{equation} \label{eq:jaynes_cummings}
    H=i\sqrt{\frac{\gamma}{\Delta t}}(b a^\dagger-b^\dagger a),
\end{equation}
where $\gamma$ is the coupling strength and $\Delta t$ is some coarse-grained time-scale in the collision model (see Eq.~(14, 16, 17) in \cite{PhysRevResearch.2.043070}), $b$ and $a$ are the lowering operators of the cavity field and the transmission line, respectively.

The original Hamiltonian in Eq.~\eqref{eq:jaynes_cummings} then generates a unitary which we keep up to order $\Delta t$:
\begin{equation*}
   U=e^{-iH\Delta t}\approx 1+\sqrt{\gamma \Delta t}(b a^\dagger-b^\dagger a)-\frac{\gamma}{2}(b a^\dagger-b^\dagger a)^2\Delta t.
\end{equation*}

The homodyne measurement readouts the quadrature basis of the probe, in-phase $I$, quadrature $Q$, or
some linear combination thereof, and can be
implemented by a variety of devices. In our physical experiments, we use JPAs. For our analysis, we will measure in the $I$ quadrature, in which we construct
the quadrature operator $R = a+a^\dagger$. Measuring in this basis, the output is a continuous variable $r$ with associated Kraus operators~\cite{livingston_continuous_2021}
\begin{equation*}
\begin{split}
     &\Omega_r = \braket{r|U|0}\\
     &=\braket{r|0}+\braket{r|1}\sqrt{\gamma \Delta t} b-\frac{\gamma}{2}\Delta t\left(\braket{r|0}b^\dagger b+\braket{r|2}\sqrt{2}b^2\right)\\
     &= \braket{r|0} \left[1+r\sqrt{\gamma\Delta t}b-\frac{\gamma}{2}\Delta t (b^\dagger b-(r^2-1)b^2)\right],
\end{split}
\end{equation*}
where $\braket{r|0}=(2\pi)^{-1/4}\exp(-r^2/4)=\sqrt{P_0(r)}$ is the probes's ground state in the position basis and $P_0(r)$ is the probability of measuring $r$ when the probe is in the ground state. In the last line, we have used the Hermite polynomials to express the harmonic oscillator's first and second excited states in terms of its ground state. 

We determine the probability of
measuring a particular outcome $r$ as
\begin{equation*}
\begin{split}
&p_r=\braket{\Omega_r^\dagger\Omega_r}_\rho\\
&=P_0(r)
\left[
1+r\sqrt{\gamma\Delta t}\braket{b+b^\dagger}_\rho+\gamma\Delta t (r^2-1)\braket{b^\dagger b}_\rho
\right],
\end{split}
\end{equation*}
where the average is taken over the states $\rho$ of the cavity field coupled to the transmons~\cite{Gambetta_Blais_Boissonneault_Houck_Schuster_Girvin_2008}.

If we approximate $r$ as a Gaussian variable, we then want to determine the mean and variance of this:
\begin{equation*}
    \begin{split}
        &\braket{r}_\rho=\int_{-\infty}^\infty r p_r dr = \sqrt{\gamma\Delta t}\braket{b+b^\dagger}_\rho,\\
        &\braket{r^2}_\rho = \int_{-\infty}^\infty r^2 p_r dr=1.
    \end{split}
\end{equation*}
Let $\Delta W$ be drawn from a Gaussian distribution with variance $\Delta t$. The statistics of the measurement record of $r$ can be reproduced by
\begin{equation} \label{eq:r_measurement_records}
     r\sqrt{\Delta t}=\sqrt{\gamma}\braket{b+b^\dagger}_\rho\Delta t+\Delta W.
\end{equation}
The voltage operator to be measured will be of the form
\begin{equation*}
    \hat{V} \propto \frac{a+a^\dagger}{\sqrt{\Delta t}},
\end{equation*}
resulting in a classical voltage
\begin{equation*}
    V=A\frac{r}{\sqrt{\Delta t}},
\end{equation*}
where $A$ is a constant scaling factor in units of $\text{V}\cdot s^{1/2}$ characterising the physical noise power in a certain bandwidth. Using Eq.~\eqref{eq:r_measurement_records}, the measured voltage $V$, which is written in terms of
\begin{equation}\label{eq:voltage}
    V\Delta t=A\left(\sqrt{\gamma}\braket{b+b^\dagger}_\rho\Delta t+\Delta W\right),
\end{equation}
has variance that scales as $\Delta t^{-1}$. The state of the transmons can be inferred from the homodyne measurement voltage in Eq.~\eqref{eq:voltage} \cite{Gambetta_Blais_Boissonneault_Houck_Schuster_Girvin_2008}.

\begin{figure}
 \centering
\includegraphics[scale=0.9]{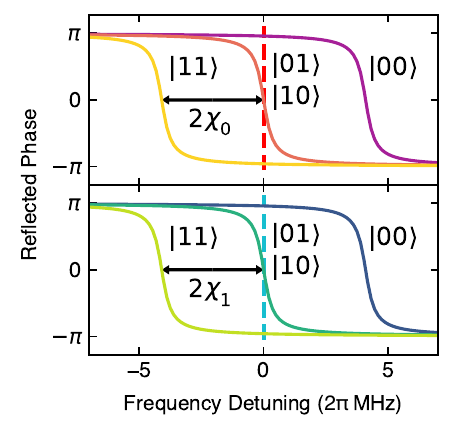}
 \caption{The reflected phase response of different two-qubit states. $\chi_0/2\pi\approx\chi_1/2\pi\approx \SI{-2}{\mega\hertz}$, where $\chi_0$ and $\chi_1$ correspond to the two resonators for the two parity measurements. We set the probing frequency to be at the center of the odd-parity resonance. Figure adapted from Fig.~1 in Ref.~\cite{Livingston_2021}.}
 \label{fig:phase_response}
\end{figure}

To implement a single parity measurement on two qubits, we dispersively couple two qubits to the same readout resonator. We tune the qubits to have the same dispersive coupling to the resonator so that the states $\ket{01}$ and $\ket{10}$ are indistinguishable on the $I$-$Q$ plane. By making the dispersive shift $\chi$ much larger than the linewidth $\kappa$ of the resonator, we can make the reflected phase of $\ket{00}$ (close to $\pi$) and $\ket{11}$ (close to $-\pi$) overlap with one another, making them indistinguishable as well. The reflected phase response is shown in Fig.~\ref{fig:phase_response}. Altogether we implement a full parity measurement of odd excitations vs. even excitations by measuring the $I$ quadrature. In our experiment, we implement two of these full parity measurements -- one between qubits $1$ and $2$ and the other between qubits $2$ and $3$ \cite{Livingston_2021}.

\section{Quantum Annealing Simulations} \label{app:annealing}
We adopt the jump/no-jump method for bit-flip errors. In this method, gradual decoherence due to the third term in Eq.~\eqref{eq:SME} is described as the average effect of bit-flip errors $X_q$ occurring at random times. At a
finite time interval $[t, t+\Delta t]$, a bit-flip error $X_q$ occurs with probability $\gamma_q \Delta t$. If this error occurs, the quantum state jumps from $\rho_t$ to $\rho_{t+\Delta t}=X_q\rho_t X_q$. Otherwise, the quantum state continuously evolves without environmental decoherence. On averaging over many instances of the bit-flip errors, the
jump/no-jump approach reduces to the open quantum system model, where errors continuously change the mixed system state $\rho(t)$.

In simulating the coherent evolution, we use the first-order Magnus expansion~\cite{blanes_pedagogical_2010} of the annealing Hamiltonian $H(t)$ in Eq.~\eqref{eq:annealing_hamiltonian} at every finite time interval $[t, t+\Delta t]$, $\tilde{U}_t = \exp{\left[-i H(t')\Delta t\right]}$
where $t'=t+\Delta t/2$, such that the quantum state evolves as $\rho_{t+\Delta t}=\tilde{U}_t \rho_t \tilde{U}_t^\dagger$. 

We average over $10,000$ quantum trajectories obtained through the above-mentioned steps to simulate the ensemble density states $\rho_t$.

\section{RNN Hidden State Size v.s. Performance} \label{app:rnn_hidden_performance}
It is desirable to limit the size of the RNN to achieve sufficiently low computational latency in real-time experiments. We present the performance in state tracking in quantum memory as described in Sec.~\ref{sec:state_tracking} for the LSTM and GRU architectures with different hidden sizes in Tab.~\ref{tab:rnn_hidden_size}. In examining the performance, we see that although we used LSTM with a hidden size $32$ in our simulated experiments, it is possible to shrink the size of the network to $16$ without harming the performance. We note that a smaller hidden size means smaller matrix-vector multiplications in computing the model, which then requires fewer memory resources in practice. The possible simplification is also suggested by the fact that the learning curves with a hidden size of $16$ is very similar to that with a hidden size of $32$, as shown in Fig.~\ref{fig:learning_curves}. Additionally, it is viable to use the GRU architecture to achieve the same performance. These results suggest that the RNN-based model may have a simpler structure and an even faster computation speed in real-time implementation on programmables like FPGAs.

We note that the size of the RNN can be further reduced, if assuming a fixed initial state so that the input to the RNN shown in Eq.~\eqref{eq:rnn_input} can be replaced by $x=[I_{1, t}, I_{2, t}]^T$.

\section{Resonator Transients} \label{app:transients}
The resonator transients are manifested from the varying SNR before the qubit-state-dependent coherent states $|\alpha_{\zeta\eta}(t)\rangle$ of the microwave field in the cavity reach their steady states when the resonator linewidth $\kappa$ is small, where $\zeta,\eta\in\{e, g\}$ and $e/g$ denotes the excited/ground state. The complex field amplitude $\langle\hat{a}\rangle_{\zeta\eta}=\alpha_{\zeta\eta}$ given that the qubits are in state $\zeta\eta$ satisfies~\cite{Gambetta_Blais_Boissonneault_Houck_Schuster_Girvin_2008, Multi-qubit, blais_circuit_2021}
\begin{equation}\label{eq:lagvine}
\begin{cases}
\dot{\alpha}_{ee}(t)=-i\varepsilon-i(\delta_r+2\chi)\alpha_{ee}(t)-\frac{\kappa}{2}\alpha_{ee}(t),\\
\dot{\alpha}_{gg}(t)=-i\varepsilon-i(\delta_r-2\chi)\alpha_{gg}(t)-\frac{\kappa}{2}\alpha_{gg}(t),
\\
\dot{\alpha}_{eg}(t)=-i\varepsilon-i\delta_r\alpha_{eg}(t)-\frac{\kappa}{2}\alpha_{eg}(t),\\
\dot{\alpha}_{ge}(t)=-i\varepsilon-i\delta_r\alpha_{ge}(t)-\frac{\kappa}{2}\alpha_{ge}(t),
\end{cases}
\end{equation}
where $\varepsilon$ is the amplitude of the driving tone, $\chi$ is the dispersive shift and $\delta_r=\omega_r-\omega_d$ is the detuning of the measurement drive to the bare cavity frequency.

The steady state ($\dot{\alpha}_{\zeta\eta}=0$) solutions to the above equations are
\begin{equation*}
\begin{cases}
    \alpha_{ee/gg}=\frac{-2\varepsilon}{2(\delta_r\pm2\chi)-i\kappa},\\
    \alpha_{eg}=\alpha_{ge}=\frac{-2\varepsilon}{2\delta_r-i\kappa}
\end{cases}
\end{equation*}
with $+$ for $ee$ and $-$ for $gg$.

In our parity measurement, we probe at the shared odd excitation resonance, which is also the same as the bare cavity frequency, i.e., $\delta_r=0$. The cavity resonance when the qubits are in $\ket{11}$ is shifted from the bare cavity resonance by $2\chi/2\pi=\SI{-4}{\mega\hertz}$, while the resonance when the qubits are in $\ket{00}$ is shifted from the bare frequency by $-2\chi/2\pi=\SI{4}{\mega\hertz}$ (see Fig.~\ref{fig:phase_response}). This results in an asymmetry between the paths in phase-space leading up to the steady states when the qubit pair changes parity. 

\begin{figure}
 \centering
\includegraphics[scale=0.48]{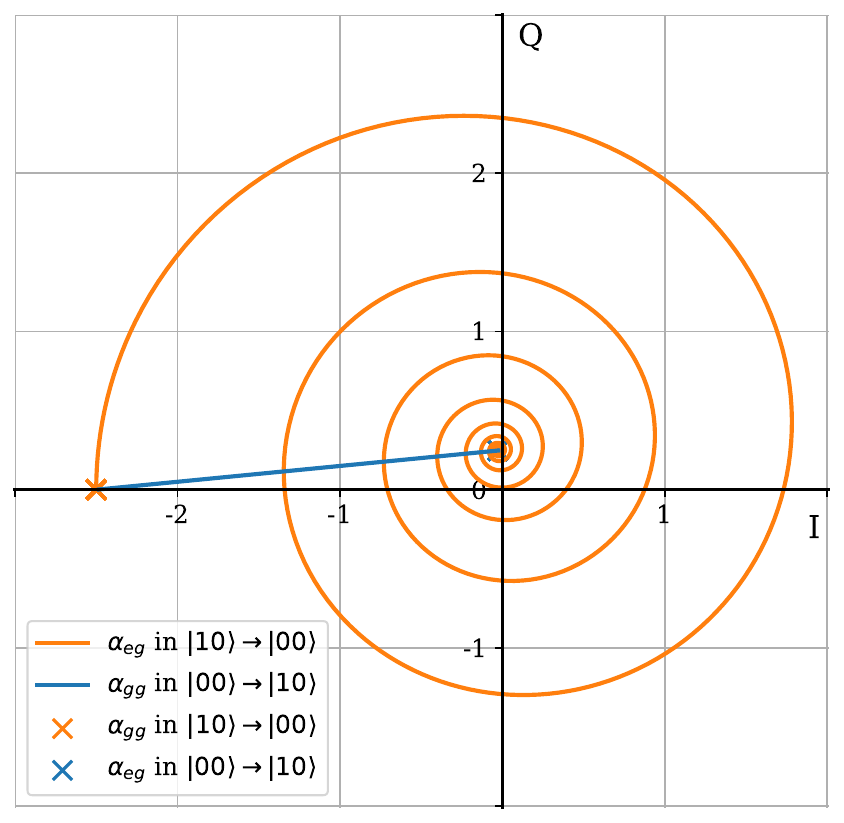}
 \caption{The pointer state paths leading up to the steady state in the phase space, with $\kappa/2\pi= \SI{800}{\kilo\hertz}$, $\chi/2\pi=\SI{-2}{\mega\hertz}$, $\delta_r=0$ and $\varepsilon$ set to $1$. When the qubit pair goes from an even parity to an odd parity, e.g., $\ket{00}\rightarrow\ket{10}$, the blue line is the path of $\alpha_{eg}(t)$ while the blue cross shows the steady state of $\alpha_{gg}$, obtained from Eq.~\eqref{eq:excite_path}. When the qubit pair goes from an odd parity to an even parity, e.g., $\ket{10}\rightarrow\ket{00}$, the orange spiral curve is the path of $\alpha_{gg}$ while the orange cross shows the steady state of $\alpha_{eg}$, obtained from Eq.~\eqref{eq:decay_path}.}
 \label{fig:spiral}
\end{figure}

When the qubits go from an even-parity state to an odd-parity state, e.g., $\ket{00}\rightarrow\ket{10}$, solving $\dot{\alpha}_{eg}(t)$ in Eq.~\eqref{eq:lagvine} with the initial coherent state at $\alpha_{gg}$ yields the path $\alpha_{eg}(t)$ specified by
\begin{equation} \label{eq:excite_path}
    \begin{cases}
    \alpha_{gg}(t) = \alpha_{gg}\\
    \alpha_{eg}(t) = \left(\alpha_{gg}+\frac{2i\varepsilon}{\kappa+2i\delta_r}\right)e^{-i\delta_r t-\frac{\kappa}{2}t}-\frac{2i\varepsilon}{\kappa+2i\delta_r}.
    \end{cases}
\end{equation}
When the qubits go from an odd-parity state to an even-parity state, e.g., $\ket{10}\rightarrow\ket{00}$, solving $\dot{\alpha}_{gg}(t)$ in Eq.~\eqref{eq:lagvine} with the initial coherent state at $\alpha_{gg}$ yields the path $\alpha_{gg}(t)$ specified by
\begin{equation} \label{eq:decay_path}
    \begin{cases}
    \alpha_{gg}(t) = \left(\alpha_{eg}+\frac{2i\varepsilon}{\kappa+2i(\delta_r-2\chi)}\right)e^{-i(\delta_r-2\chi)t-\frac{\kappa}{2}t}-\frac{2i\varepsilon}{\kappa+2i(\delta_r-2\chi)}\\
    \alpha_{eg}(t) = \alpha_{eg}.
    \end{cases}
\end{equation}
These paths are shown in Fig.~\ref{fig:spiral}. Strictly speaking, the two sets of solutions apply when there are no dynamics apart from the dispersive measurements. 

The measurement strength is defined as \cite{Korotkov_2016, Gambetta_Blais_Boissonneault_Houck_Schuster_Girvin_2008}
\begin{equation*}
\begin{split}
    \Gamma(t)&=\frac{1}{2}\kappa\abs{\alpha_{gg}(t)-\alpha_{eg}(t)}^2,
\end{split}
\end{equation*}
which scales the separation of the two parity signal means under constant noise variance (see Eq.~\eqref{eq:signal_expression}). In the odd-to-even parity transition, the path in phase-space leading up to the steady states forms a tighter spiral as the ratio $\abs{\chi/\kappa}$ gets larger. A tighter spiral translates to a more oscillatory $\Gamma(t)$, thus leading to a more oscillatory signal mean \cite{blais_circuit_2021}.

\begin{figure}
 \centering
\includegraphics[scale=0.6]{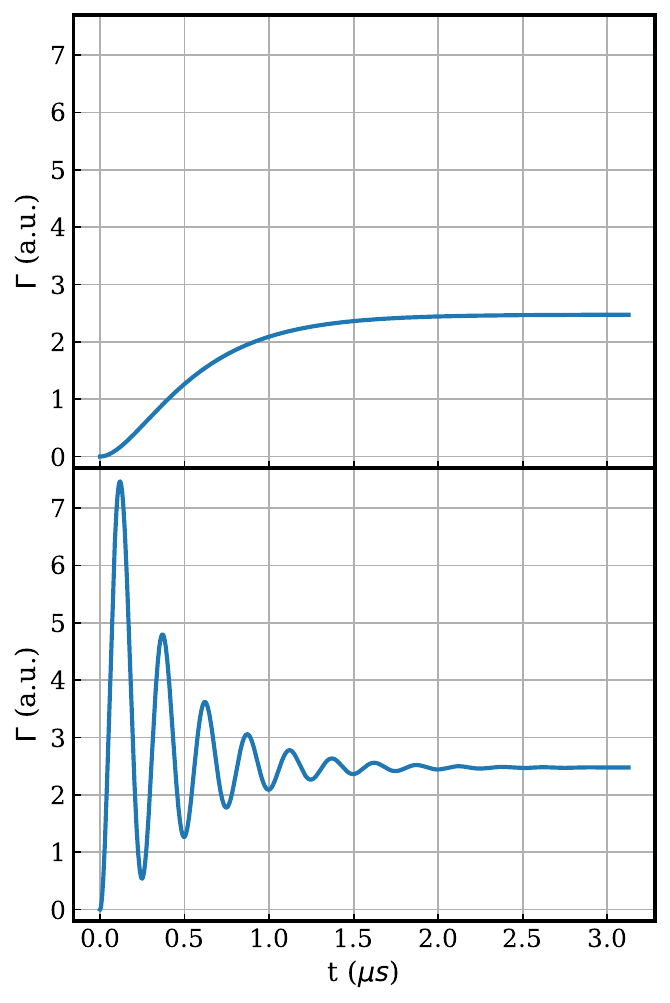}
 \caption{The measurement rate $\Gamma(t)$ on a pair of qubits with a bit-flip transition at $t=0$, with $\kappa/2\pi= \SI{800}{\kilo\hertz}$, $\chi/2\pi=\SI{-2}{\mega\hertz}$, $\delta_r=0$ and $\varepsilon$ set to $1$. The upper figure corresponds to the qubit pair transitioning from an even parity to an odd parity, obtained from Eq.~\eqref{eq:excite_path}. The lower figure corresponds to the the qubit pair transitioning from an odd parity to an even parity, obtained from Eq.~\eqref{eq:decay_path}.}
 \label{fig:ring}
\end{figure}

Shown in Fig.~\ref{fig:ring}, the ring-up transient without clear oscillations is manifested in the measurement strength corresponding to the even-to-odd parity transition in Eq.~\eqref{eq:excite_path}, whereas the ring-down transient with oscillations is manifested in the measurement strength corresponding to the odd-to-even parity transition in Eq.~\eqref{eq:decay_path}. They show good agreement with experimental observations, such as those in Fig.~\ref{fig:signal_example}.

\section{Population of States Subject to Amplitude Damping or Bit-flips}\label{app:population_curve}
We recall that the population of the excited states $P_\text{exc}$ is the ensemble population of the states that are at most one bit-flip away from the fully excited state $\ket{111}$, i.e., $P_\text{exc}=P(\ket{111})+P(\ket{110})+P(\ket{101})+P(\ket{011})=P_7+P_6+P_5+P_3$. 

Under $T_1$ decay at zero temperature, the transition matrix evolving the states for time $T$ is $J'(T)=\exp(Q'T)$, where $Q'$ is defined as,
\begin{equation}\label{eq:t1_Q_transition}
Q'=\gamma \begin{bmatrix} 0 & 0 & 0 & 0 & 0 & 0 & 0 & 0\\
                   1 & -1 & 0 & 0 & 0 & 0 & 0 & 0\\
                   1 & 0 & -1 & 0 & 0 & 0 & 0 & 0\\
                   0 & 1 & 1 & -2 & 0 & 0 & 0 & 0\\
                   1 & 0 & 0 & 0 & -1 & 0 & 0 & 0\\
                   0 & 1 & 0 & 0 & 1 & -2 & 0 & 0\\
                   0 & 0 & 1 & 0 & 1 & 0 & -2 & 0\\
                   0 & 0 & 0 & 1 & 0 & 1 & 1 & -3\\
 \end{bmatrix}.
\end{equation}
The state probabilities under the Markov chain are given by $P(T)=J'(T)P(0)$, which yields 
\begin{equation*}
    P_\text{exc}(T) = \left(3e^{\gamma T}-2\right)e^{-3\gamma T}.
\end{equation*}

Under only bit-flip errors $X_q$, the transition matrix evolving the states for time $T$ is $J(T)=\exp(QT)$, where $Q$ is defined in Eq.~\eqref{eq:Q_bitflip}. The resultant population of excited states is
\begin{equation*}
    P_\text{exc}(T) = e^{-3\gamma T}\cosh^2(\gamma T)\left[3\sinh(\gamma T)+\cosh(\gamma T)\right].
\end{equation*}

\section{Performance Comparison between the Double Threshold Method and the Double Threshold Boxcar Filter}\label{app:boxcar}

\begin{figure}[h]
 \centering
\includegraphics[scale=0.45]{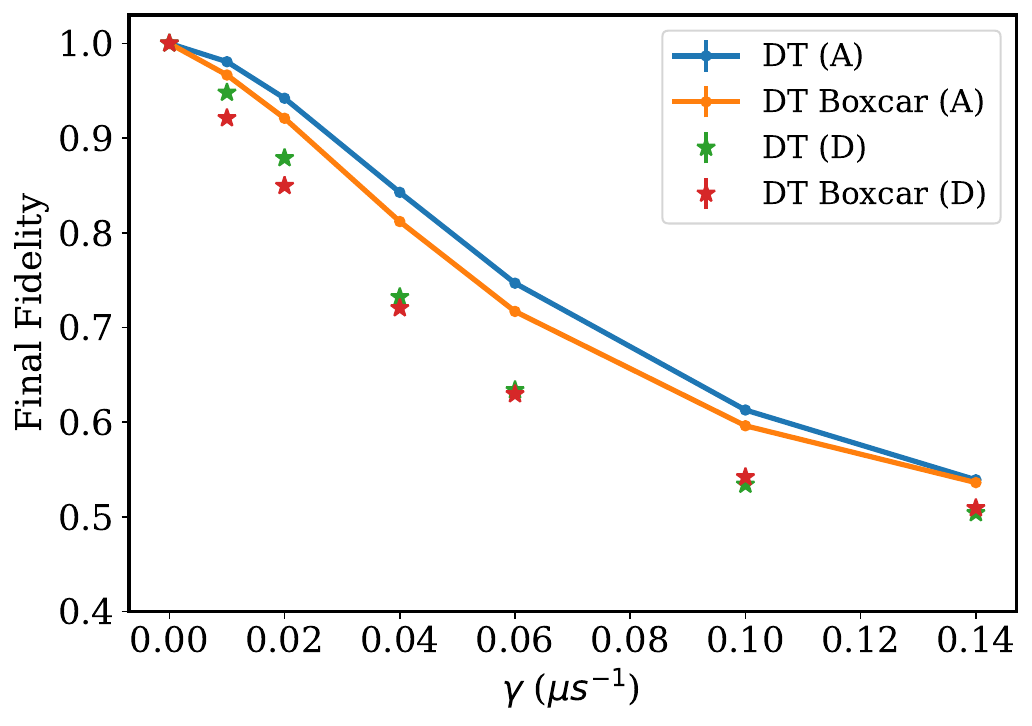}
 \caption{The final fidelity with respect to the initial state $\ket{000}$ in Schemes A and D with the double threshold exponential filter (DT) in \cite{atalaya_continuous_2021}, and the double threshold boxcar filter (DT Boxcar) in \cite{mohseninia_2020}, as a function of single qubit bit-flip rate $\gamma$ at an operation time $T=\SI{20}{\micro\second}$ with a measurement strength $\Gamma_m=\SI{4.7}{\per\micro\second}$. Each data point is averaged over $30,000$ quantum trajectories.}
 \label{fig:boxcar}
\end{figure}

The double threshold boxcar filter in \cite{mohseninia_2020} employs a boxcar averaging of the measurement signals and two thresholds, one fixed at zero and the other at a variable position above zero. We compare the performance of this against the double threshold model (with exponential filter and two variable thresholds) from \cite{atalaya_continuous_2021} that was used in this work, by running the state tracking task as described in Sec.~\ref{sec:state_tracking} on Schemes A and D, as shown in Fig.~\ref{fig:boxcar}. The double threshold method outperforms double threshold boxcar in both schemes at relatively low error rates.

\end{appendix}
\end{document}